\documentclass[lettersize,journal]{IEEEtran}
\usepackage{amsmath,amsfonts}
\usepackage{algorithmic}
\usepackage{algorithm}
\usepackage{array}
\usepackage[caption=false,font=normalsize,labelfont=sf,textfont=sf]{subfig}
\usepackage{textcomp}
\usepackage{stfloats}
\usepackage{url}
\usepackage{verbatim}
\usepackage{graphicx}
\usepackage{cite}
\usepackage{siunitx}
\usepackage[]{collab}
\usepackage{enumitem}
\usepackage{booktabs}
\usepackage{multirow}
\usepackage{hyperref}

\newcommand\alias{\textsc{LogPrism}\xspace} 

\collabAuthor{zb}{teal}{Zhuangbin}
\collabAuthor {ly}{purple}{Liuyang}
\collabAuthor {km}{orange}{Kaiming}

\newsavebox{\obsbox} 
\newenvironment{summarybox}{%
  \par\vspace{6pt}\noindent 
  \setlength{\fboxrule}{0.2pt}%
  \setlength{\fboxsep}{3pt}%
  \begin{lrbox}{\obsbox}%
    \begin{minipage}{\dimexpr\linewidth-2\fboxsep-2\fboxrule\relax}
    \vspace{1pt}%
}{%
    \end{minipage}%
  \end{lrbox}%
  \fcolorbox{black}{white!90!black}{\usebox{\obsbox}}%
  \par\vspace{6pt} 
}

\begin{document}

\title{\alias: Unifying Structure and Variable Encoding for Effective Log Compression}  

\author{\IEEEauthorblockN{Yang Liu, Kaiming Zhang, Zhuangbin Chen$^{\dag}$\thanks{$^{\dag}$Zhuangbin Chen is the corresponding author.}, Zibin Zheng}

\IEEEauthorblockA{School of Software Engineering, Sun Yat-sen University, Zhuhai, China\\ \{liuy2355, zhangkm7\}@mail2.sysu.edu.cn, \{chenzhb36, zhzibin\}@mail.sysu.edu.cn}}

\maketitle

\begin{abstract}  
In the field of log compression, the prevailing ``\textit{parse-then-compress}'' paradigm fundamentally limits effectiveness by treating log parsing and compression as isolated objectives. While parsers prioritize semantic accuracy (i.e., event identification), they often obscure deep correlations between static templates and dynamic variables that are critical for storage efficiency. In this paper, we investigate this misalignment through a comprehensive empirical study and propose \alias, a framework that bridges the gap via \textit{unified redundancy encoding}. Rather than relying on a rigid pre-parsing step, \alias dynamically integrates structural extraction with variable encoding by constructing a \textit{Unified Redundancy Tree (URT)}. This hierarchical approach effectively mines ``structure+variable'' co-occurrence patterns, capturing deep contextual redundancies while accelerating processing through pre-emptive pattern encoding. Extensive experiments on 16 benchmark datasets confirm that \alias establishes a new state-of-the-art. It achieves the highest compression ratio on 14 datasets, surpassing existing baselines by margins of 6.12\% to 83.34\%, while delivering superior throughput at 29.87 MB/s (1.68$\times$$\sim$43.04$\times$ faster than competitors). Moreover, when configured in single-archive mode to maximize global pattern discovery,
\alias boosts its compression ratio by 273.27\%, outperforming the best baseline by 19.39\% with a 2.62$\times$ speed advantage.
\end{abstract}


\section{Introduction}
\label{sec:introduction}



Software systems generate logs to record runtime events, errors, and operational states, which are indispensable for system maintenance~\cite{DBLP:conf/issre/HeZHL16,DBLP:journals/corr/abs-2107-05908, DBLP:conf/ccs/Du0ZS17, DBLP:conf/icml/XuHFPJ10,DBLP:conf/asplos/YuanMXTZP10,DBLP:journals/infsof/YuanLSL20,DBLP:journals/corr/abs-2509-26463,DBLP:conf/sigsoft/Zhou0X0JLXH19} and performance optimization~\cite{DBLP:journals/csur/HeHCYSL21, DBLP:conf/sigsoft/ChenKLZZXZYSXDG20,DBLP:conf/osdi/ZhaoRLYS16}.
However, the sheer volume of log data poses significant storage challenges~\cite{DBLP:conf/usenix/DingZLZLFZX15,DBLP:journals/pacmse/ChenPZ25}.
Modern large-scale systems can produce terabytes or even petabytes of logs daily~\cite{DBLP:conf/osdi/RodriguesLY21, DBLP:conf/osdi/ChowMFPW14, DBLP:conf/usenix/YangPO18}.
Thus, efficient log compression has become a critical concern for managing storage costs while preserving the analytical value of historical log data~\cite{DBLP:conf/osdi/RodriguesLY21,DBLP:conf/osdi/WangGR0ZWFC024}.

Logs typically follow a ``template+variables'' pattern, where static strings are interleaved with dynamic runtime parameters.
General-purpose compression algorithms like gzip~\cite{DBLP:journals/rfc/rfc1951} and LZMA~\cite{DBLP:journals/tit/ZivL77} fail to exploit the specific characteristics of log data, often resulting in suboptimal compression ratios~\cite{DBLP:journals/ese/YaoLSH20}.
To address this, recent research has focused on log-specific compression methods that leverage the inherent semi-structured nature of logs. By separating the template and variable components, parser-based compressors can replace repetitive templates with compact identifiers and compress variable streams with specialized encoding techniques, achieving significantly higher efficiency.

Despite these advancements, existing approaches face a fundamental limitation, i.e., the decoupling of log parsing~\cite{DBLP:conf/icse/ZhuHLHXZL19,DBLP:conf/issta/JiangL00HGCZL24} and compression.
Current workflows treat parsers as a black-box pre-processor, which aim to maximize semantic accuracy (i.e., identifying the correct templates) without regard for downstream compression effectiveness.
To quantify the impact of this misalignment, we conduct a comprehensive empirical study evaluating four state-of-the-art compressors and nine parsers. Our findings reveal that semantically accurate parsers may produce templates that undermine compression performance, such as over-generalized templates that offload excessive entropy to the variable stream, or over-fitted templates that incur substantial dictionary overhead. Furthermore, the conventional ``parse-then-compress'' pipeline creates a boundary between the log structure and parameters, preventing the exploitation of deeper redundancies. Specifically, it ignores \textit{template-variable correlations}, where specific variable values are strongly tied to a template and could be encoded as part of the structure, and \textit{inter-variable correlations}, where variables within a single log entry co-occur in predictable patterns. By treating these components in isolation, existing methods fail to mine and encode these high-value aggregate patterns.

To overcome these limitations, we propose \alias, a log compression framework that unifies structural extraction and variable encoding. Instead of relying on a pre-defined parsing stage, \alias constructs a \textit{Unified Redundancy Tree (URT)} that dynamically models both log structure and variable correlations in an integrated representation. Our approach employs a hierarchical redundancy mining strategy that progressively distills log data through three stages. First, we extract stable log tokens to construct a structural tree, establishing a compact skeleton for subsequent analysis. Second, we extend the skeleton by building variable subtrees to mine frequent ``structure+variable'' co-occurrence patterns, effectively bridging the gap between templates and parameters. Finally, we execute residual data processing to efficiently handle the remaining high-entropy ``long-tail'' variables using a specialized sorting and stream normalization pipeline.

This design maximizes compression ratios by capturing deep contextual redundancies while simultaneously accelerating processing speed. By filtering out the majority of high-frequency patterns in the early stages, \alias drastically reduces the computational load on the final, expensive residual processing stage. We further enhance scalability through a parallel-aware architecture that supports fine-grained concurrency. 
Extensive experiments on 16 datasets from LogHub~\cite{DBLP:conf/icse/ZhuHLHXZL19} demonstrate that \alias sets a new state-of-the-art for both effectiveness and efficiency. It achieves the highest compression ratio on 14 datasets (surpassing existing baselines by 6.12\%$\sim$83.34\%) while delivering the fastest processing speeds (1.68$\times$$\sim$ 43.04$\times$ faster than competitors). Moreover, we address a critical trade-off of baselines, which rely on chunking dataset to enable parallelism at the cost of limited global pattern discovery. When configured in non-chunked mode, \alias boosts its compression ratio by 273.27\%, outperforming the leading compressor, Denum, by 19.39\% in the same mode. Particularly, \alias's speed in this computationally intensive mode (15.77 MB/s) remains comparable to Denum's default chunked performance (17.83 MB/s).

In summary, this paper makes the following contributions:

\begin{itemize}[noitemsep,leftmargin=5.5mm]
    \item We conduct the first comprehensive empirical study to quantify the impact of log parsers on compression, revealing the critical misalignment between parsing accuracy and compression efficiency.

    \item We propose the concept of unified redundancy encoding, a paradigm shift that co-designs structural extraction and variable encoding to exploit deep ``structure+variable'' correlations. Based on this concept, we design and implement \alias, a high-performance log compression framework featuring a unified redundancy tree and a parallel-aware architecture.

    \item We perform an extensive evaluation demonstrating that \alias significantly outperforms existing state-of-the-art methods in both compression ratio and speed.
\end{itemize}
\section{Background and Motivation}
\label{sec:background}

\subsection{Parser-based Log Compression}

\begin{figure}[t]
    \centering
    \includegraphics[width=\columnwidth]{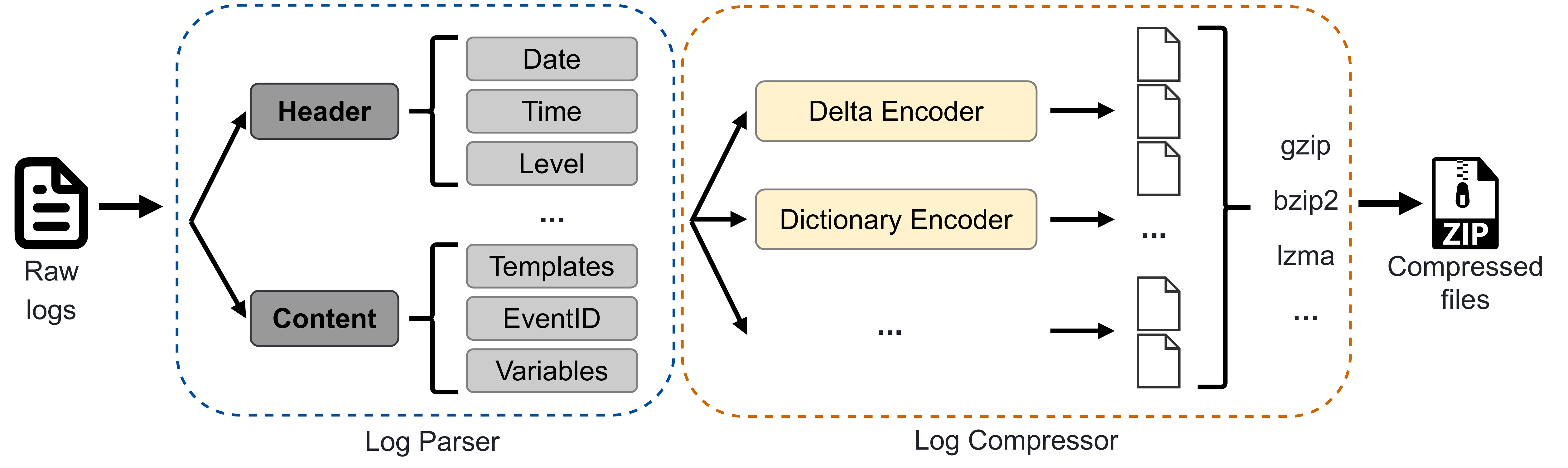}
    \caption{The General Workflow of Parser-based Log Compression}
    \label{fig:parser-based_log_compression}
\end{figure}

In the realm of log compression, parser-based methods represent the predominant approach.
They leverage the inherent semi-structured nature of logs, typically composed of static format strings and dynamic parameters. The process begins with a critical prerequisite step, i.e., \textit{log parsing}. As illustrated in Fig.~\ref{fig:parser-based_log_compression}, parsers like Drain~\cite{DBLP:conf/icws/HeZZL17} are employed to decompose raw log messages, separating structured headers (e.g., timestamps, log levels) from the free-form body, and then segment the body into an invariant event template and its corresponding variables. This structured representation enables the core compression mechanism, where repetitive template text is replaced with compact identifiers (stored only once in a dictionary) and the extracted parameters are aggregated for subsequent encoding using specialized compression techniques.
The resulting output, which is a highly regular stream of template IDs and parameter arrays with significantly reduced entropy, is then processed by general-purpose algorithms like LZMA or gzip to eliminate any remaining statistical redundancy.
Several representative log compressors~\cite{DBLP:conf/kbse/LiuZHHZL19,DBLP:conf/fast/WeiZWLZCSZ21,DBLP:conf/icse/LiZL024,DBLP:conf/kbse/YuW0H24} exemplify different design philosophies within this framework.

The performance of these parser-based compressors is fundamentally tied to the quality of the log parsing stage~\cite{DBLP:conf/kbse/YuW0H24,DBLP:journals/ese/KhanSBB24}.
However, existing research mainly focuses on compression algorithms that operate on the parsed logs.
Their designs, therefore, default to an assumption of ideal parsing quality. The parser is often treated as an independent component rather than an integral part of the compression pipeline whose performance demands critical evaluation. This has led to a common practice where an off-the-shelf parser is selected based on secondary criteria like execution speed or ease of integration.
However, there is no one-size-fits-all parser. Log parsers employ a variety of heuristics and algorithms, and their accuracy can vary significantly across different datasets with diverse log formats.
Any errors introduced during parsing~\cite{DBLP:conf/icse/Khan0BB22, DBLP:conf/issta/JiangL00HGCZL24, DBLP:conf/icse/LiLCSHLZ23, DBLP:conf/sigsoft/FuYXLLZY22} will inevitably propagate and undermine the effectiveness of the downstream compression algorithms.
Despite this critical dependency, the interplay between parsing accuracy and compression efficiency remains largely unexplored. To systematically evaluate the impact of different log parsers on the performance of log compression, and to further reveal the critical role of parsing in the compression pipeline, we conduct a comprehensive empirical study in this paper.



\subsection{Revisiting Log Parsers in Compression Pipelines}
\label{sec:empirical_study}


In this section, we perform an empirical study to investigate how the choice of log parsers quantitatively affects the performance of existing log compressors.
We collect a set of representative log parsers and parser-based log compressors, and perform controlled experiments across multiple benchmark datasets that are widely used in log analysis domain.

\subsubsection{Experiment Setup}

We introduce the experiment setup of our study as follows:


\textbf{Log Parser Selection:}
We select nine representative log parsers, i.e., Drain~\cite{DBLP:conf/icws/HeZZL17}, AEL~\cite{DBLP:conf/qsic/JiangHFH08}, IPLoM~\cite{DBLP:conf/kdd/MakanjuZM09}, LFA~\cite{DBLP:conf/msr/NagappanV10}, LogSig~\cite{DBLP:conf/cikm/TangLP11}, MoLFI~\cite{DBLP:conf/iwpc/MessaoudiPBBS18}, SHISO~\cite{DBLP:conf/IEEEscc/Mizutani13}, Spell~\cite{DBLP:conf/icdm/Du016} and the parser implemented in LogReducer~\cite{DBLP:conf/fast/WeiZWLZCSZ21}.
By covering this broad spectrum of algorithmic strategies, our study can generate heterogeneous template sets and provide a robust evaluation of their impact on downstream compression performance.
These parsers span seven distinct methodological families, from simple heuristics to complex data-driven models.
For instance, AEL represents the heuristic approach that uses lightweight, generic rules to distinguish static text from dynamic parameters, requiring minimal domain knowledge.
In contrast, Drain and LogReducer employ fixed-depth parsing trees, where root-to-leaf paths define templates, enabling efficient online processing.
Other parsers frame the task as a data mining problem. LFA applies frequent pattern mining and IPLoM iteratively partitions log messages based on token position and cardinality.
Clustering-based parsers like LogSig and SHISO identify templates by computing pairwise message similarity and treat each resulting cluster's centroid as a template.
Our selection also includes other novel strategies.
Spell identifies templates based on the longest common subsequence algorithm, making it particularly effective for streaming data.
MoLFI leverages evolutionary algorithms to iteratively evolve optimal template sets via genetic operations.

\textbf{Log Compressor Selection}:
To select representative log compressors, we review existing solutions~\cite{DBLP:conf/kbse/LiuZHHZL19,DBLP:journals/pacmse/ChenPZ25, DBLP:conf/fast/WeiZWLZCSZ21, DBLP:conf/icse/YuCLWZDZ23, DBLP:conf/icse/LiZL024, DBLP:conf/kbse/YuW0H24, DBLP:conf/IEEEscc/Chen24, DBLP:conf/osdi/RodriguesLY21, DBLP:conf/eurosys/WeiZ00ZSWJ23} in the literature and apply two filtering criteria.
First, we exclude online compressors~\cite{DBLP:journals/pacmse/ChenPZ25,DBLP:conf/icse/YuCLWZDZ23}.
Our study focuses on offline lossless compression for long-term storage, which prioritizes maximizing storage efficiency across the entire dataset.
In contrast, online methods optimize for reducing real-time transmission overhead, often operating under tight resource constraints and with the possibility of lossy compression~\cite{DBLP:conf/nsdi/ZhangXAVM23,DBLP:conf/icse/YuCLWZDZ23,DBLP:conf/IEEEcloud/ChenJSLZ24}.
Second, we prioritize selecting the state-of-the-art approaches that represent the most effective methodologies in the field.
This process yields four representative compressors: Logzip~\cite{DBLP:conf/kbse/LiuZHHZL19}, LogReducer~\cite{DBLP:conf/fast/WeiZWLZCSZ21}, LogShrink~\cite{DBLP:conf/icse/LiZL024}, and Denum~\cite{DBLP:conf/kbse/YuW0H24}.
While Denum is primarily a number-centric approach designed to bypass traditional template extraction, we include it because it 
relies on parser-based strategies like LogShrink to compress its non-numeric log content.
Consequently, how the input text is parsed or tokenized remains a critical factor in Denum's overall performance.

These approaches feature a different design in log structure extraction and encoding strategies, providing a diverse testbed for our study. Logzip, as a pioneer in parser-based compression, employs iterative clustering to uncover latent structures, transforming logs into a hierarchical representation with template identification and parameter mapping. LogReducer advances this idea by focusing on inter-parameter correlations, utilizing an elastic numeric encoding scheme to dynamically select compact bit representations, complemented by delta encoding for sequential data. Similarly, LogShrink optimizes parameter storage by reorganizing data into columnar formats, grouping homogeneous types to create low-entropy streams ideal for dictionary encoding~\cite{DBLP:journals/tit/ZivL78}. Finally, Denum introduces a number-centric approach that targets numerical data, using regular expressions to isolate numeric strings for categorization and differential encoding to minimize redundancy.

\begin{table}[t]
    \centering
    \caption{Detailed Statistics of Benchmark Log Datasets}
    \label{tab:dataset_stats}
    \begin{tabular}{l l r r}
        \toprule
        \textbf{System Type} & \textbf{Dataset} & \textbf{File Size} & \textbf{\# Lines} \\
        \midrule
        \multirow{5}{*}{Distributed Systems} 
        & HDFS & 1.47 GB & 11,175,629 \\
        & Hadoop & 48.61 MB & 394,308 \\
        & Spark & 2.75 GB & 33,236,604 \\
        & Zookeeper & 9.95 MB & 74,380 \\
        & OpenStack & 58.61 MB & 207,820 \\
        \midrule
        \multirow{3}{*}{Supercomputers} 
        & BGL & 708.76 MB & 4,747,963 \\
        & HPC & 32.00 MB & 433,489 \\
        & Thunderbird & 29.60 GB & 211,212,192 \\
        \midrule
        \multirow{3}{*}{Operating Systems} 
        & Windows & 26.09 GB & 114,608,388 \\
        & Linux & 2.25 MB & 25,567 \\
        & Mac & 16.09 MB & 117,283 \\
        \midrule
        \multirow{2}{*}{Mobile Systems} 
        & Android & 183.37 MB & 1,555,005 \\
        & HealthApp & 22.44 MB & 253,395 \\
        \midrule
        \multirow{2}{*}{Server Applications} 
        & Apache & 4.90 MB & 56,481 \\
        & OpenSSH & 70.02 MB & 655,146 \\
        \midrule
        \multirow{1}{*}{Standalone Software} 
        & Proxifier & 2.42 MB & 21,329 \\
        \bottomrule
    \end{tabular}
\end{table}

\textbf{Dataset Selection:} To ensure generalizability and robustness, our evaluation utilizes the widely recognized LogHub benchmark~\cite{DBLP:conf/icse/ZhuHLHXZL19}. As detailed in Table~\ref{tab:dataset_stats}, this collection comprises 16 datasets spanning a broad spectrum of system types, including large-scale distributed systems (e.g., HDFS, Spark), supercomputers (BGL, Thunderbird), server applications (Apache, OpenSSH), operating systems (e.g., Linux, Windows), mobile platforms (Android, HealthApp), and standalone software (Proxifier). In total, the full benchmark contains over 77 GB of raw logs and approximately 378 million log entries. However, our preliminary experiments revealed that the Android and Windows datasets could not be processed by some parsers within a reasonable time budget due to their inherent complexity. Consequently, we exclude these two datasets from our empirical study. Our final evaluation is conducted on the remaining 14 datasets, which collectively account for over 50 GB of data and 262 million log entries, spanning diverse log formats and structural complexities.

\textbf{Evaluation Metrics:} To quantify how different log parsers impact the effectiveness of downstream compression, we employ the Compression Ratio (CR), a standard metric in data compression. It is defined as the ratio of the original log file size to the compressed file size. A higher value indicates better compression performance.

\begin{equation*}
    Compression~Ratio~(CR) = \frac{Original~Log~Size}{Compressed~File~Size}
\end{equation*}

\subsubsection{Experiment Workflow}

We design a controlled workflow for the study, which is structured into three phases, i.e., \textit{log parsing}, \textit{intermediate normalization}, and \textit{log compression}.

In the initial phase, we apply the nine selected log parsers to generate structured template collections for each dataset. During this process, we observe significant inconsistencies in parser outputs regarding wildcard notation, delimiter usage, and file formats. For example, some parsers use \texttt{<*>} to denote variables, while others use markers like \texttt{spec} or simply omit the variable tags.
Such heterogeneity creates format incompatibilities that can prevent downstream compressors from correctly recognizing templates.
To eliminate these discrepancies, we perform intermediate normalization in the second phase.
This standardizes all parser outputs into a single canonical format, while preserving semantic content.
Specifically, we unify all wildcard symbols, align metadata structures to a uniform schema, and convert outputs to a consistent file format.
This ensures that compression performance differences reflect parsing quality rather than formatting artifacts.
In the final phase, we evaluate the four selected compressors using these normalized templates. This presents a technical challenge, as most off-the-shelf log compressors are monolithic systems with tightly coupled parsing and compression modules. To address this, we systematically refactor their source code, decoupling data ingestion from the core compression logic. We develop dedicated input interfaces that allow us to inject our normalized templates, creating a controlled environment to test each compressor against templates from all nine parsers.

\subsubsection{Experimental Results}

\begin{figure*}[t]
    \centering
    \includegraphics[width=1\textwidth]{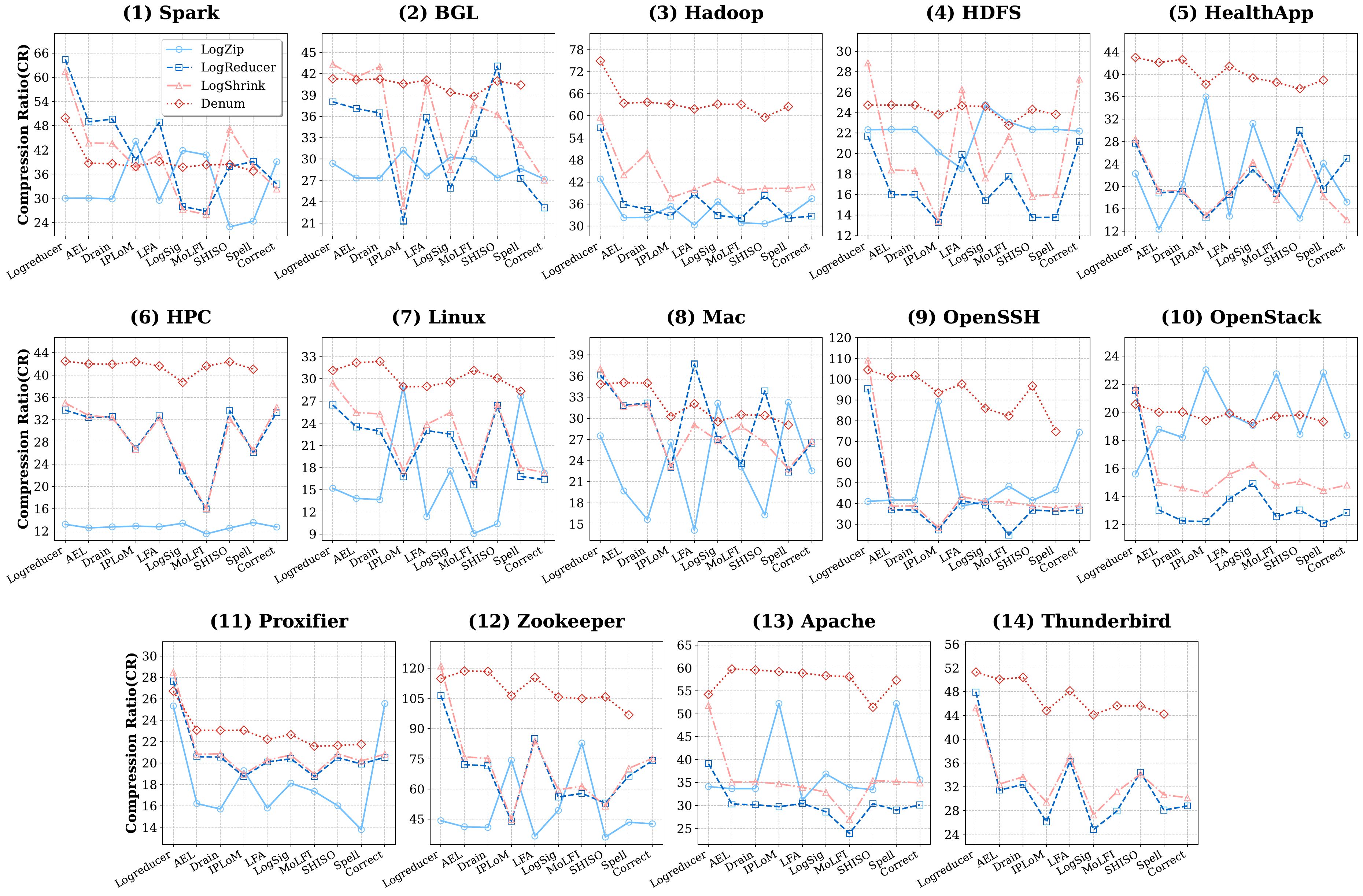}
    \caption{Impact of Log Parser Selection on Downstream Compression Ratio across Different Datasets}
    \label{fig:empirical_study_results}
\end{figure*}

Fig.~\ref{fig:empirical_study_results} presents the compression ratio achieved by four log compressors when supplied with templates generated by nine different parsers. Our analysis reveals three key observations that fundamentally challenge the current parser-based paradigm for log compression.

\textbf{Observation 1: Dramatic Performance Variance Induced by Parser Selection.}
For any given compressor, the choice of parser induces dramatic variance in the final compression ratio, often surpassing the inherent performance differences between the compressors themselves.
Three compressors (LogZip, LogReducer, and LogShrink) exhibit particularly strong sensitivity.
For example, on Zookeeper, LogShrink achieves a CR of 121.06 when using templates from LogReducer's parser, but this plummets to just 44.81 with the IPLoM parser's templates.
Denum, which is not fully parser-based, also demonstrate notable performance fluctuations, confirming that the handling of non-numeric text remains critical.
Such results provide clear evidence that parser selection plays a critical, yet previously overlooked, role in compression efficiency.

\textbf{Observation 2: Default parsers are not always optimal.}
Counter-intuitively, our experiments reveal that compressors can achieve higher compression ratios using external parsers rather than their built-in ones.
On BGL, for instance, LogReducer achieves a CR of 38.04 when using its own default parser, but this increases to 43.08 when paired with the SHISO parser.
A similar trend is observed on the HealthApp dataset, where SHISO again outperforms LogReducer's native parsing logic (29.98 vs. 27.71). These findings suggest that the tight coupling of specific parsers with compressors in current designs can be suboptimal, preventing the compression algorithms from realizing their full potential on diverse datasets.

\textbf{Observation 3: Parsing Accuracy Does Not Guarantee Compression Efficiency.}
Our results demonstrate that there is no universally optimal parser capable of consistently maximizing compression efficiency across all datasets and compressors. While Drain generally performs well, it is significantly outperformed by SHISO and LFA on Thunderbird. Conversely, SHISO excels on BGL but yields suboptimal results on HDFS and OpenStack.
Even the ground-truth templates (labeled ``Correct'') are frequently outperformed by heuristic parsers.
This variance indicates that a parser's effectiveness for compression is not intrinsic but depends on the complex interplay between data characteristics and the compressor's encoding strategy. The root cause of this volatility lies in a fundamental misalignment of objectives: parsers prioritize classification accuracy to maximize event clustering correctness, whereas compressors prioritize storage efficiency. Parsers emphasize semantic accuracy without considering the storage cost of the resulting templates and parameters.
This leads to two typical inefficiencies that degrade compression performance despite high parsing accuracy:



\begin{itemize}[noitemsep,leftmargin=5.5mm]
    \item \textit{Over-generalization (Coarse-grained Templates)}: Parsers prioritizing high matching rates with fewer templates often introduce excessive wildcards. On HDFS, for instance, LFA generates only 42 templates but with a total of 372 wildcards (8.8 per template on average). While this may yield high parsing accuracy, it offloads complexity to the parameter stream. The compressor is forced to process large, noisy parameter sets that should have been static template text, severely degrading the compression ratio.
    

    \item \textit{Over-fitting (Fine-grained Templates):} Conversely, overly sensitive clustering thresholds can partition semantically similar logs into numerous distinct templates. For example, some parsers produce over 28,000 templates for the HealthApp dataset, which contains only around 150 actual event types. This ``template explosion'' exponentially increases the dictionary size, creating a direct storage overhead that undermines compression efficiency despite high formal parsing precision.
    
\end{itemize}

In summary, our empirical study confirms that log parsing is not a mere preprocessing step but a dominant factor in the ultimate efficiency of log compression.
The fundamental principle of exploiting log structural redundancy is sound,
which enables parser-based approaches consistently and significantly outperforming general-purpose algorithms like gzip or LZMA. However, our findings reveal a critical flaw in its common decoupled implementation.
We argue that unlocking optimal log storage performance requires a paradigm shift where structural extraction is not a prerequisite for compression but an integral, co-designed part of it.


\begin{figure}[t]
    \centering
    \includegraphics[width=\columnwidth]{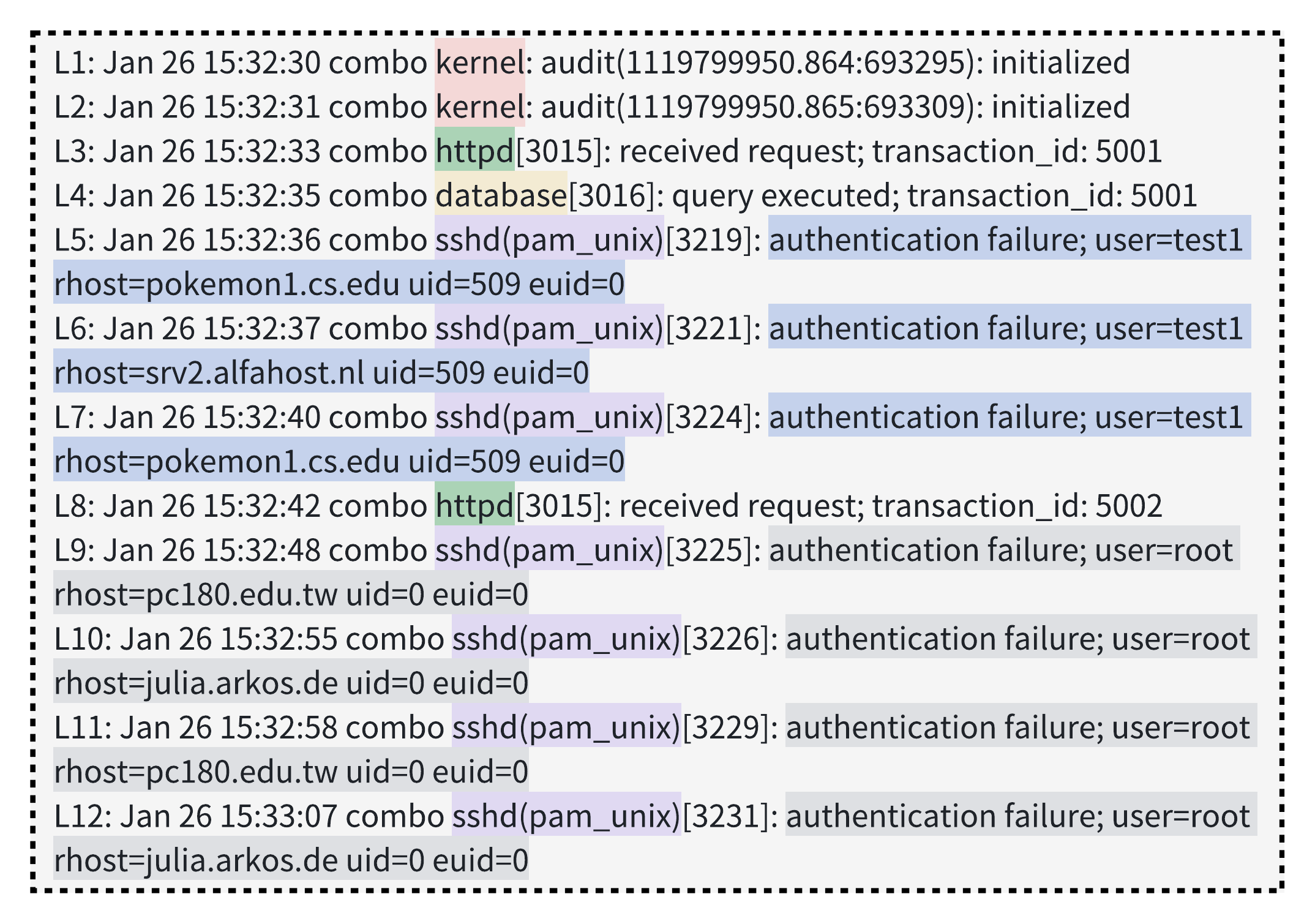}
    \caption{A 12-line Log Snippet as a Running Example}
    \label{fig:running_example}
\end{figure}

\subsection{A Motivating Example for Unified Redundancy Encoding}
\label{sec:motivating_example}

In this section, we illustrate the core concept of our approach based on the logs in Fig.~\ref{fig:running_example}.
The idea is to move beyond the rigid dichotomy of ``static strings vs. dynamic variables'' by modeling log tokens (whether traditionally considered a static string or a variable) uniformly and encode them in aggregate based on their co-occurrence and dependencies.
To this end, we propose a paradigm shift from the traditional \textit{parse-then-compress} workflow and introduce a new methodology called \textit{unified redundancy encoding}.


Consider the \texttt{sshd} log entries in Fig.~\ref{fig:running_example} (L5-L7 and L9-L12).
Since variable \texttt{user}, \texttt{rhost}, and \texttt{uid} take different values, a conventional parser may generate a log template like \texttt{"...authentication failure; user=} \texttt{<*> rhost=<*> uid=<*> euid=0"}.
When compressing these logs, e.g., L5, existing compressors would first store an identifier for this template and then independently encode the variable values (\texttt{test1}, \texttt{pokemon1.cs.edu}, and \texttt{509}).
In this process, two types of correlations are ignored. The first is \textit{template-variable correlation}. For pure log compression purposes,
the template ID could include certain variables if they are strongly tied to the template body, eliminating separate parameter processing. 
Although parsers may occasionally inline frequent variables (e.g., \texttt{euid=0} in Fig.~\ref{fig:running_example}) into the template, this behavior is inconsistent and not measurable.
The second is \textit{inter-variable correlation} within each log entry, which allows for the collective processing of variables. For instance, user \texttt{test1} deterministically maps to uid \texttt{509}, and user \texttt{root} maps to uid \texttt{0}.
This differs from prior correlation mining methods~\cite{DBLP:conf/icse/LiZL024,DBLP:conf/kbse/YuW0H24} that mainly model relationships across log entries (e.g., incremental user IDs).

Our method is more holistic and context-aware. It recognizes that the value set $\{\texttt{test1}, \texttt{pokemon1.cs.edu}, \texttt{509}\}$ (L5 and L7) is a frequent pattern occurring within the context of the aforementioned template.
Therefore, our method aggregates both parts as \texttt{"...authentication fail- ure; user=test1 rhost=pokemon1.cs.edu uid= 509 euid=0"} and encodes it with a single ID.
Similar frequent patterns include $\{\texttt{root}, \texttt{pc180.edu.tw}, \texttt{0}\}$ (L9 and L11) and $\{\texttt{root}, \texttt{julia.arkos.de}, \texttt{0}\}$ (L10 and L12), while only \texttt{srv2.alfahost.nl} (L6) needs to be handled independently.
The advantage of this unified redundancy encoding is significant: whereas existing approaches require storing one template ID plus three separate variable IDs, we need only a single ID to represent the entire, highly correlated token sequence.
This principle of co-designing structural extraction and pattern encoding enables us to discover and exploit deep contextual redundancies in log data.

A straightforward way to implement this idea is to build a prefix tree (i.e., a trie) over the sequences of log tokens and assign IDs to frequent paths, thereby identifying recurring log (sub)sequences. However, this naive construction presents two fundamental limitations.
First, high-cardinality fields induce many branches, causing the number of nodes to grow with the product of distinct values per position. This results in prohibitive storage consumption and increased lookup latency due to structural inefficiency.
Second, a trie only encodes prefixes.
If an infrequent token appears early, the match stops and the rest of the log line is no longer compressible, even when the suffix is highly regular (e.g., \texttt{uid=509} \texttt{euid=0} after a rare \texttt{rhost} in L6).
These constraints motivate our design of \alias, an effective log compressor which moves beyond pure prefix matching and supports compact, gap-tolerant aggregation of frequent token sequences.
\section{Methodology}
\label{sec:methodology}

\begin{figure*}[t]
    \centering
    \includegraphics[width=1\textwidth]{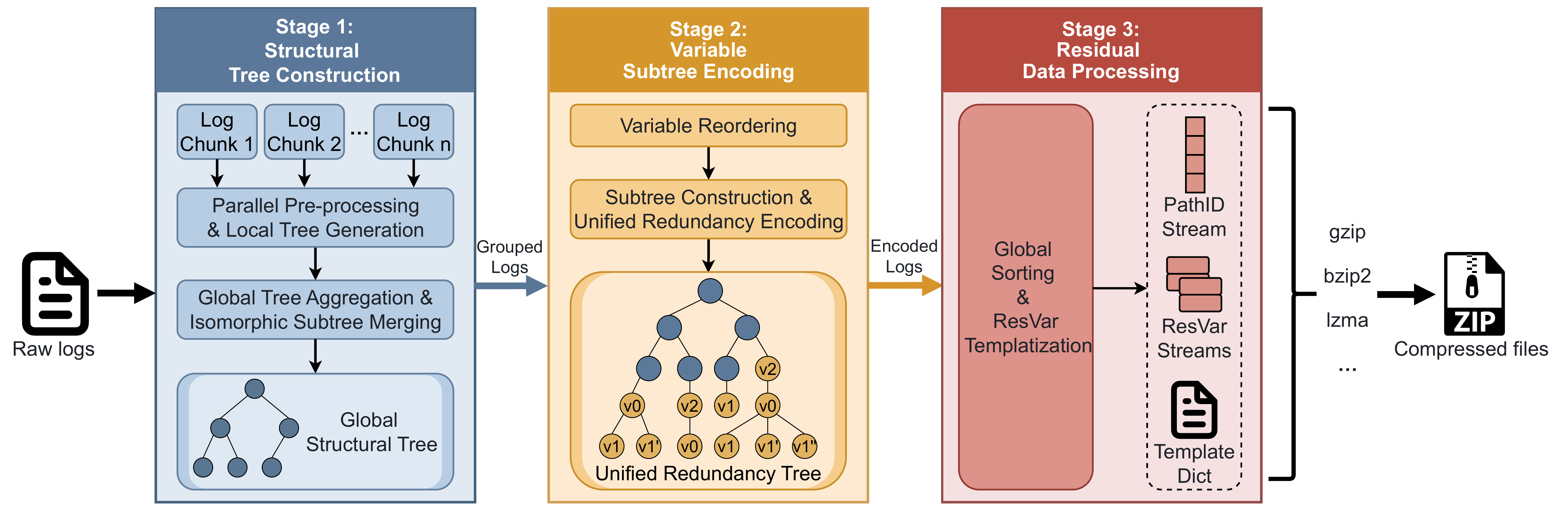}
    \caption{An Overview of the \alias Compression Framework}
    \label{fig:overview}
\end{figure*}

In this section, we introduce \alias, our log compression framework that constructs a \textit{Unified Redundancy Tree (URT)} to jointly model log structure and variable correlations. Since a naive prefix tree over raw logs would suffer from combinatorial branch explosion, \alias employs a hierarchical redundancy mining strategy to build a structurally efficient URT.
We progressively integrate log tokens into the URT based on their frequency and stability through three stages, i.e., \textit{Structural Tree Construction}, \textit{Variable Subtree Encoding}, and \textit{Residual Data Processing}, as shown in Fig.~\ref{fig:overview}.
The first stage builds the structural skeleton of the URT based on regular, fixed tokens.
Each path in this tree represents a group of log entries that are structurally similar, regardless of the log events they represent. 
The second stage integrates frequent variable tokens by expanding the skeleton's terminal nodes into subtrees.
Within each subtree, we mine deep correlations between the structural skeleton and specific variable values, allowing a single ID to represent complex, recurring ``structure-variable'' patterns.
The third stage handles the remaining ``long-tail'' tokens (i.e., outliers that are too rare or random) by isolating and encoding them via specialized schemes suited for high-entropy data.
The entire compression pipeline concludes by feeding all compressed data into a general-purpose compressor (e.g., LZMA) to exploit remaining byte-level redundancy.

It is essential to note that \alias's hierarchical compression pipeline is fundamentally different from the decoupled ``parse-then-compress'' workflow. \alias's ultimate goal is to unify the encoding of both structural and variable tokens into an end-to-end redundancy pattern, rather than treating them as separate entities for isolated compression.

\subsection{Structural Tree Construction}
\label{sec:structural_tree_construction}

The primary objective of this stage is to construct the foundational skeleton of the URT by extracting stable patterns from the raw logs. To ensure high throughput and a low memory footprint,
we implement this process within a parallel streaming architecture. The log dataset is partitioned into multiple chunks, each processed concurrently by a dedicated worker thread to build local structures. As depicted in Fig.~\ref{fig:overview}, this construction phase proceeds through two key steps: 1) \textit{Parallel Pre-processing and Local Tree Generation}, where logs are tokenized, filtered, and organized into local prefix trees; and 2) \textit{Global Tree Aggregation and Isomorphic Subtree Merging}, which integrates these local trees into a unified global structure while dynamically refining the topology.


\subsubsection{Parallel Pre-processing and Local Tree Generation}

The pipeline within each worker thread involves two operations, i.e., the pre-processing of the assigned log chunk to filter volatile content, and the construction of a local structural tree.

To prevent the ``branch explosion'' associated with high-cardinality data, we identify and separate two categories of rapidly-changing tokens.

\begin{figure}[t]
    \centering
    \includegraphics[width=\columnwidth]{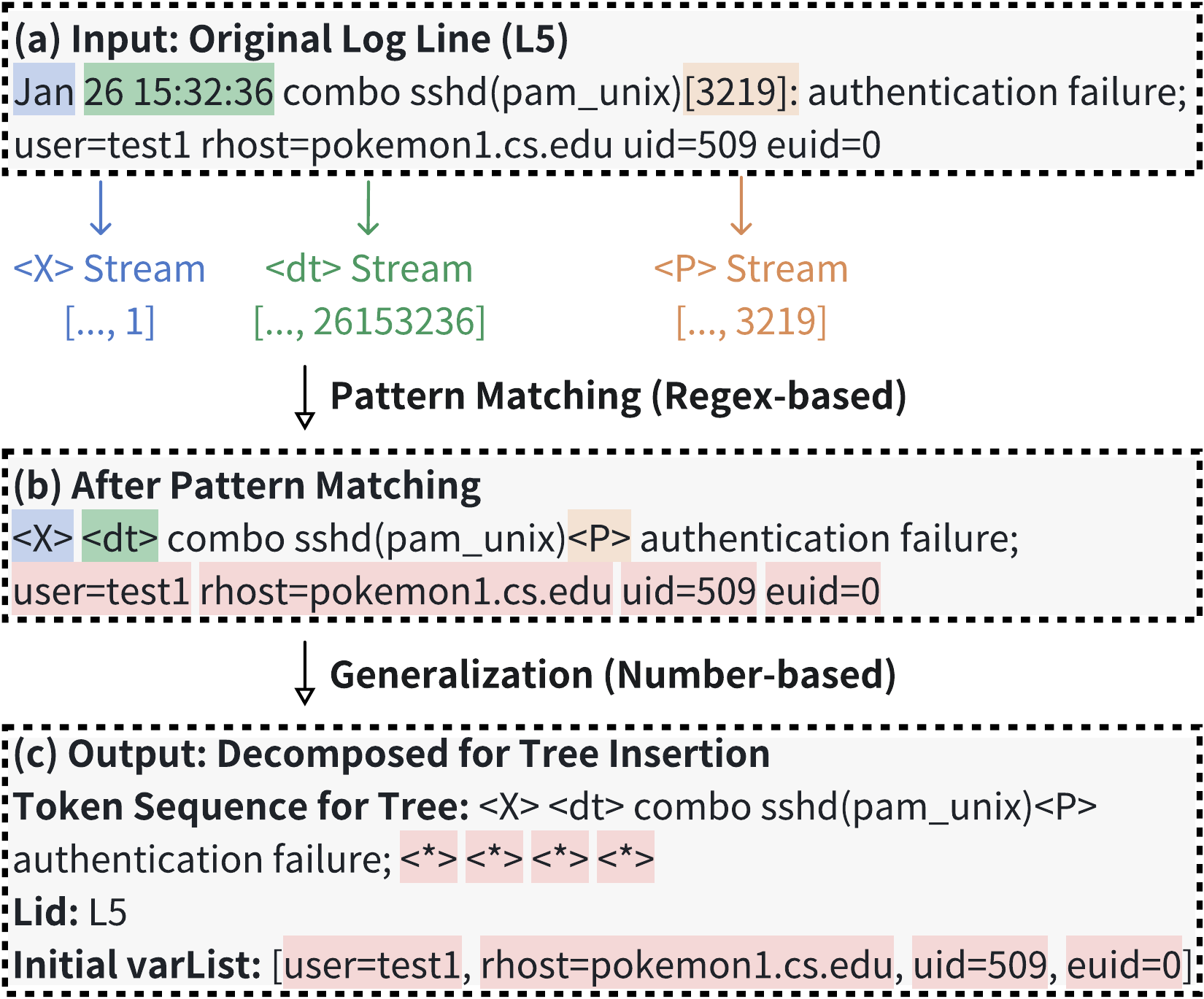}
    \caption{The Pre-processing Pipeline for Log Entry L5}
    \label{fig:preprocessing_pipeline}
\end{figure}

\begin{itemize}[noitemsep,leftmargin=5.5mm]
    \item \textit{Globally Patterned Metadata}: This category includes common header fields like dates, timestamps, and process IDs (PIDs)~\cite{DBLP:conf/icse/ZhuHLHXZL19}. These tokens, while highly variable, exhibit predictable syntactical patterns. Storing them in a string-based tree is highly inefficient. Thus, we identify them using predefined regular expressions and extract their values into separate, highly compressible columnar streams, preserving the global order of the logs. 
    As illustrated in Fig.~\ref{fig:preprocessing_pipeline}(a), these tokens are replaced with specific placeholders (e.g., $\langle X \rangle$ for month, $\langle dt \rangle$ for timestamp, $\langle P \rangle$ for PID) in raw logs.

    \item \textit{Unstructured Numeric Tokens}: 
    This category targets tokens within free-text log contents that are likely to be variables.
    Based on a simple and effective heuristic from prior research~\cite{DBLP:conf/kbse/YuW0H24}, we treat tokens containing numeric characters as potential variables.
    We apply this rule to tokens not captured by the metadata regex and uniformly replace the matched ones by the wildcard placeholder $\langle * \rangle$, as shown in Fig.~\ref{fig:preprocessing_pipeline}.
    The original values of these tokens are collected sequentially into a \texttt{varList} associated with the specific log entry.
    In this process, two types of misclassifications can happen.
    First, static tokens containing numbers (e.g., \texttt{node1}) are generalized as variables. Since \alias treats static and dynamic tokens as a holistic entity, these tokens can be re-integrated into the structure during correlation mining stage (Sec.~\ref{sec:variable_redundancy_subtree_encoding}).
    Second, string-only variables (e.g., \texttt{user=root}) are treated as static, causing potential tree fragmentation. This is explicitly addressed via isomorphic subtree merging in the subsequent step.
\end{itemize}

\begin{figure*}[t]
    \centering
    \includegraphics[width=1\textwidth]{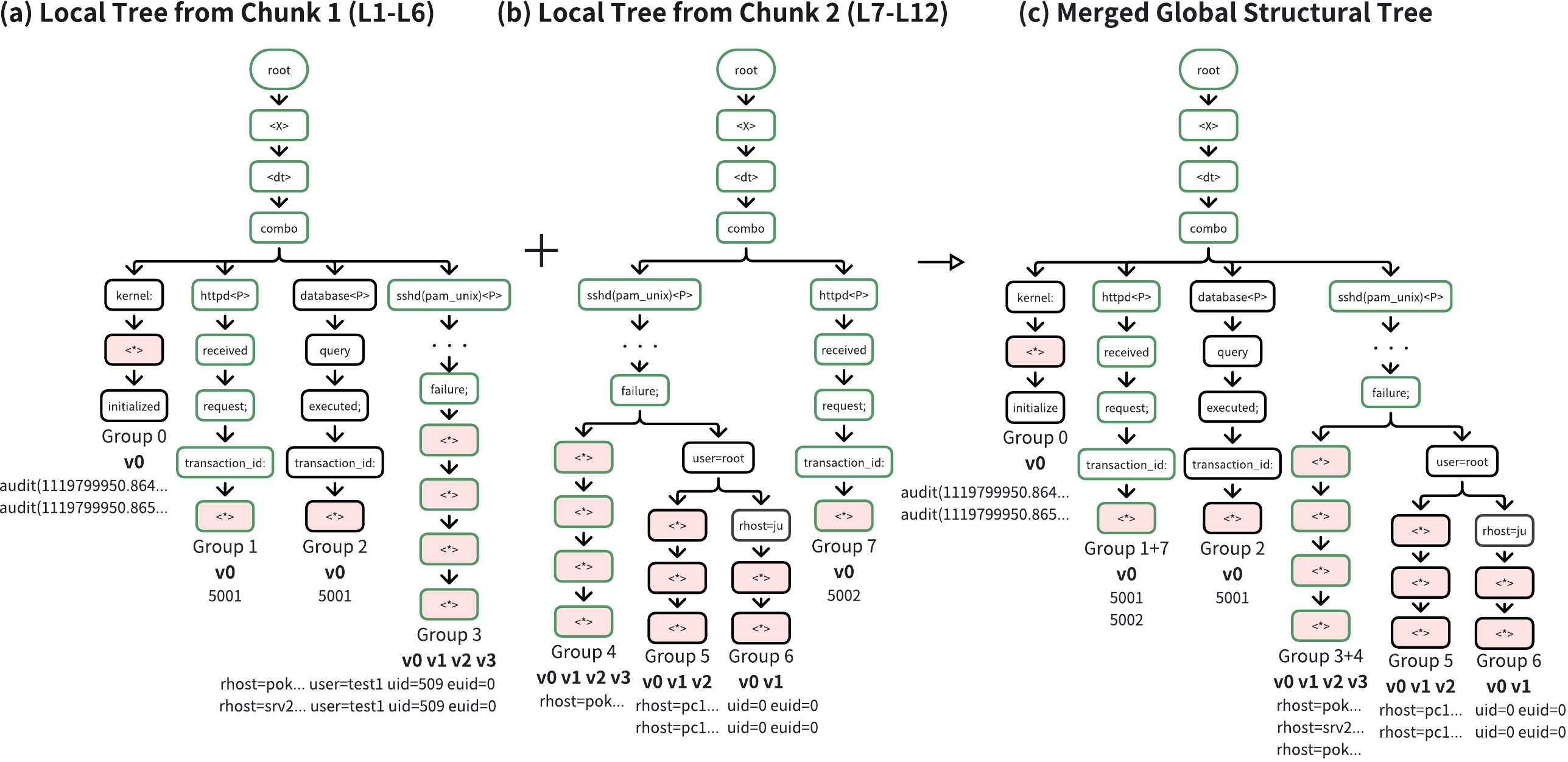}
    \caption{The Parallel Pre-processing and Streaming Merge Process of the Structural Tree}
    \label{fig:structural_tree_merge}
\end{figure*}

After pre-processing, the resulting log messages are used to construct a local prefix tree.
For example, dividing the logs from Fig.~\ref{fig:running_example} into two chunks (L1-L6 and L7-L12) yields the independent trees shown in Fig.~\ref{fig:structural_tree_merge}(a) and Fig.~\ref{fig:structural_tree_merge}(b), respectively.
In this structure, every terminal node marks the end of a log message.
Crucially, a terminal node is not necessarily a leaf node, since one log entry can be a complete prefix of another.
Each path from the root to a terminal node uniquely identifies a group of log entries sharing the same structural pattern.
To index these groups, a \texttt{Record} Object is maintained at every such terminal node, aggregating the Line IDs (\texttt{Lid}) and extracted \texttt{varList}(s) for all corresponding logs.

\subsubsection{Global Tree Aggregation and Isomorphic Subtree Merging}

Once the worker threads generate their local structures, the main thread integrates them into a single global tree and refines the topology.
We employ an on-arrival merge strategy to minimize synchronization overhead. As soon as a worker thread completes its chunk, its local prefix tree is merged into the global one.
As depicted in Fig.~\ref{fig:structural_tree_merge}(c), this merge process aggregates information along identical paths and adds new branches where structures differ (e.g., the \texttt{user=root} branch from the second chunk). Concurrently, the columnar data streams from each thread are appended to global files, preserving the original order of the log dataset.
For numerical streams like $\langle dt \rangle$,
we apply a specialized compression pipeline consisting of Delta Encoding (to store value differences), ZigZag Encoding (to efficiently represent negative numbers), and Varint Encoding (for variable-length integer representation) to generate a final compact binary format.

Although the number-based heuristic for identifying volatile fields in the first step is effective, it can potentially misclassify purely string-based variables.
For instance, \texttt{user=test1} is correctly generalized to $\langle * \rangle$ (due to \texttt{"1"}), but \texttt{user=root} would be incorrectly identified as a stable token. This inconsistency creates unnecessary branches and fragments the tree. To resolve this, we introduce a correction process named \textit{isomorphic subtree merging}.
It operates on the principle that if a ``static'' branch and a ``variable'' branch lead to subtrees that are topologically isomorphic, they serve the same structural role and should be treated uniformly as a variable.

\begin{figure}[t]
    \centering
    \includegraphics[width=\columnwidth]{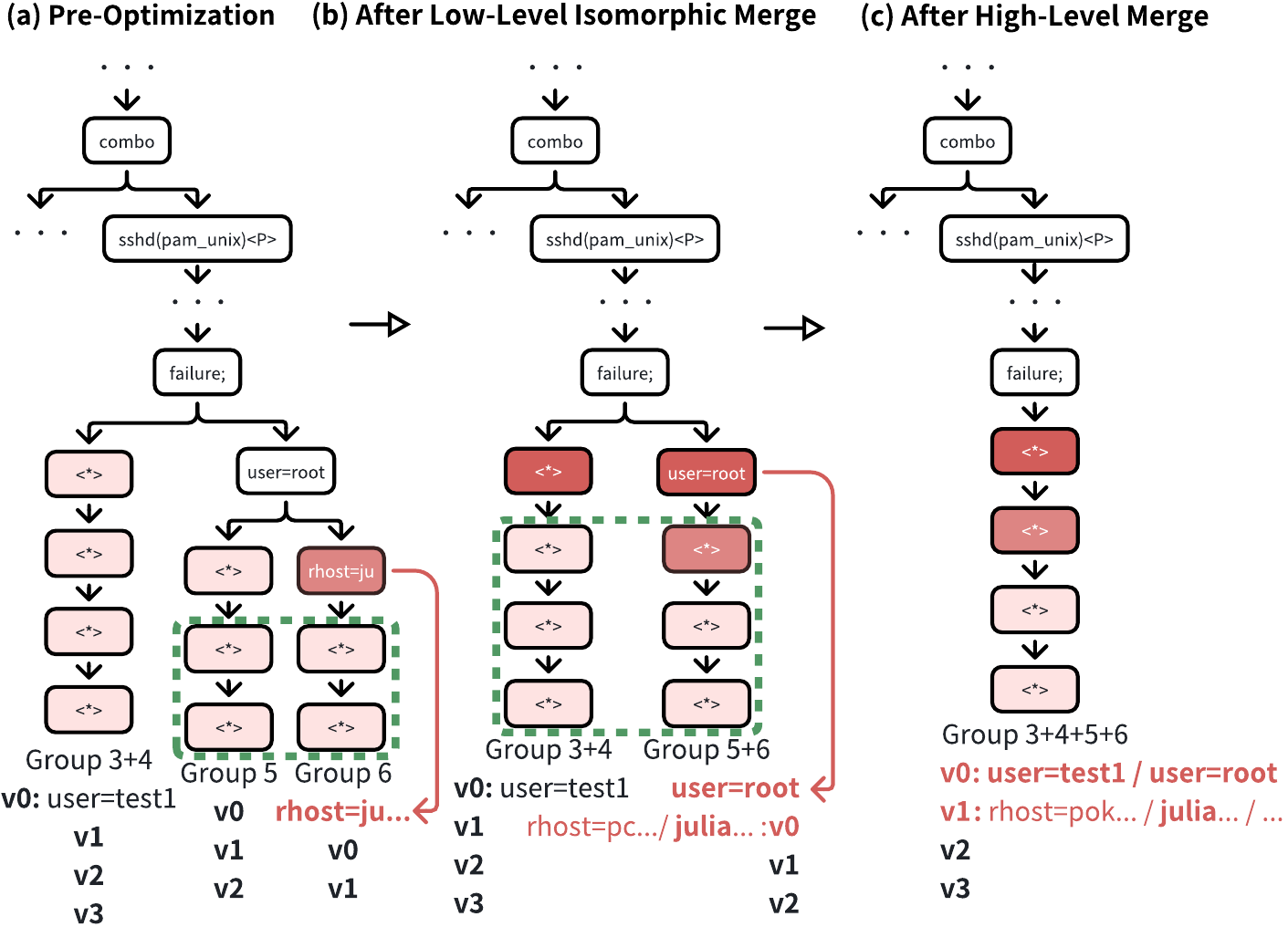}
    \caption{The Iterative, Bottom-up process of Isomorphic Subtree Merging}
    \label{fig:homomorphic_optimization}
\end{figure}

The merging process is implemented as a post-order (bottom-up) traversal of the global tree.
At each branching node, it computes a unique structural signature for every child subtree.
If a node contains a linear path, its signature is simply the concatenation of its tokens. For nodes with multiple branches, we lexicographically sort the signatures of all outgoing branches before concatenating them.
Any sibling nodes sharing identical structural signatures will be considered isomorphic, and the algorithm performs a merge operation.
Fig.~\ref{fig:homomorphic_optimization} demonstrates this iterative merging.
The traversal moves up from the leaves and, as shown in Fig.~\ref{fig:homomorphic_optimization}(a), encounters its first branching node \texttt{user=root}. It compares the child nodes $\langle * \rangle$ (which represents numeric \texttt{rhost}s) and \texttt{rhost=julia.arkos.de} by computing the structural signature for the path under each. Since both yield the same sequence of $\langle * \rangle$ nodes (as highlighted by the green dashed box), they are topologically identical. This isomorphism indicates that \texttt{rhost=julia.arkos.de} should also be a variable token (i.e., $\langle * \rangle$). The system then merges these two paths by (i) inserting the value \texttt{"rhost=julia.arkos.de"} into the corresponding position of the \texttt{varList} (the red arrow line) for all logs traversing that path, and (ii) merging the \texttt{Record} Object (i.e., \texttt{Lid}s and updated \texttt{varList}s) from the string branch into the sibling wildcard branch.
The result is the more generalized structure in Fig.~\ref{fig:homomorphic_optimization}(b).
As the traversal continues up to the next branching node, \texttt{failure;}, the system performs another similar merge as shown in Fig.~\ref{fig:homomorphic_optimization}(c).


The choice of a bottom-up traversal is critical for accurate path mergings. By processing the tree from the leaves upward, the algorithm ensures that sibling nodes sharing the longest common prefixes are evaluated for merging first. Such paths are more likely to be truly isomorphic~\cite{DBLP:conf/icws/HeZZL17}.
It also ensures that any structural inconsistencies at deeper levels are resolved first, recursively generalizing the tree into its most compact form.
In case of incorrect mergings, \alias can separate the outlier variables during the correlation mining stage (Sec.~\ref{sec:variable_redundancy_subtree_encoding}).

\begin{figure}[t]
    \centering
    \includegraphics[width=0.54\columnwidth]{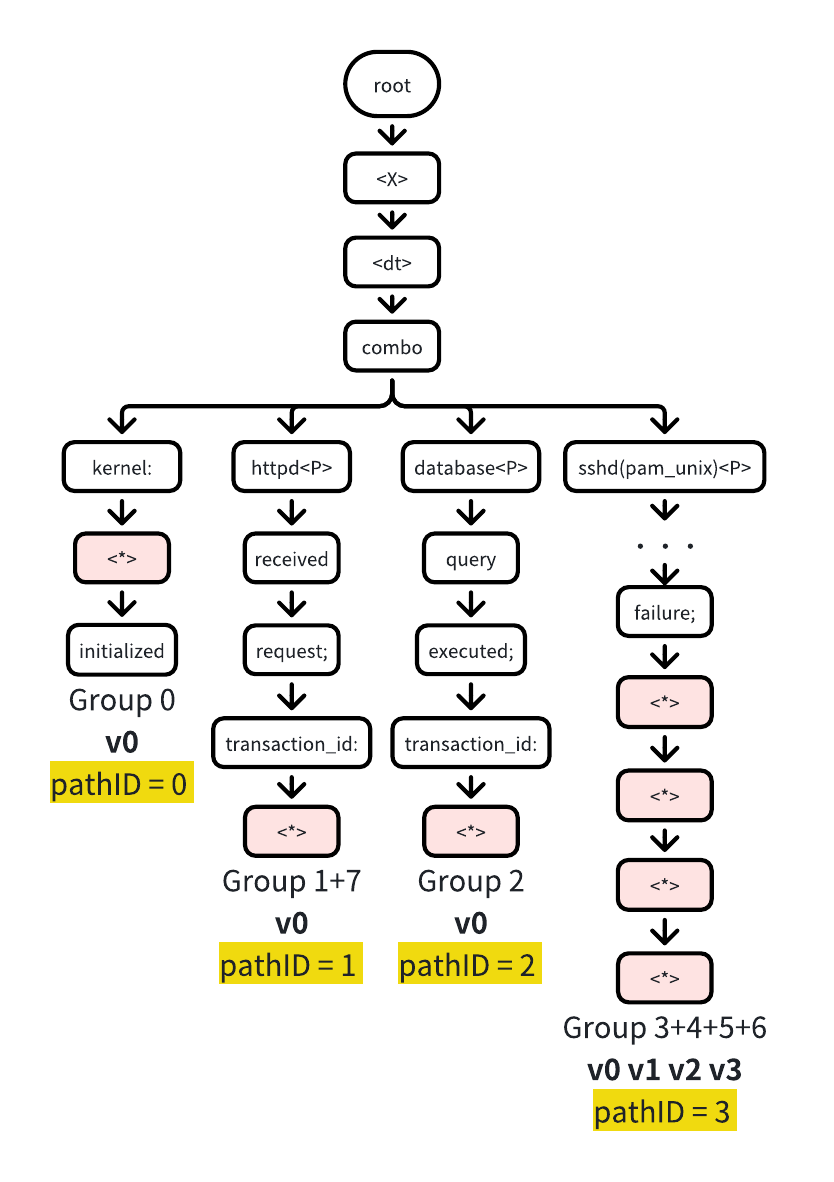}
    \caption{The Final Global Structural Tree with \texttt{pathID} Assignments}
    \label{fig:final_structural_tree}
\end{figure}

Finally, we traverse the global tree to index the structural contexts.
Every terminal node is assigned a globally unique \texttt{pathID}. As shown in Fig.~\ref{fig:final_structural_tree}, this ID acts as a \textit{structural context identifier}, representing a group of logs that share the exact same skeletal structure. For example, all \texttt{sshd} logs (L5-L7 and L9-L12) in the example are mapped to \texttt{pathID=3}.
The core role of the \texttt{pathID} is to provide clear log groupings for the next stage of variable correlation analysis, where each log is represented by its original line number (\texttt{Lid}), its structural context identifier (\texttt{pathID}), and its \texttt{varList}.


\subsection{Variable Subtree Encoding}
\label{sec:variable_redundancy_subtree_encoding}



\begin{figure}[t]
    \centering
    \includegraphics[width=1\columnwidth]{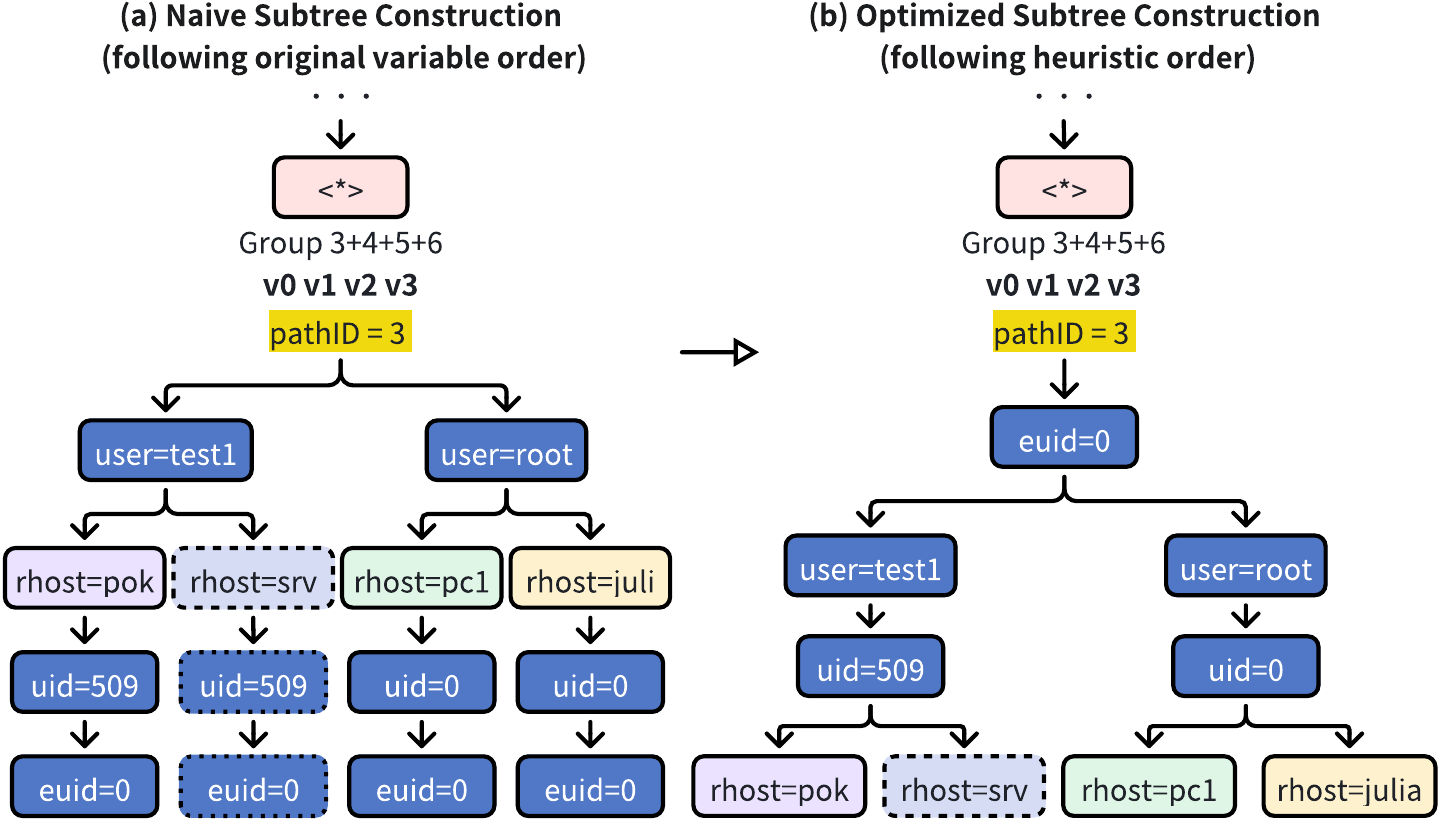}
    \caption{The Motivation and Effect of Variable Reordering}
    \label{fig:variable_reordering}
\end{figure}

The construction of the structural context tree leverages only stable tokens, resulting in a highly compact skeleton where all variable tokens are aggregated at the terminal nodes.
To capture the deep correlations between the log structure and its parameters, we can conceptually extend the URT by attaching a prefix subtree to each terminal node, using the \texttt{varList}s of the associated log entries as input paths.
By doing so, we unify the structural skeleton and variables into a single, continuous representation.
However, this presents two critical challenges.
\textit{First, a naive sequential insertion of variables inevitably triggers a secondary branch explosion.} As illustrated in Fig.~\ref{fig:variable_reordering}(a), if variables are processed in their original order (e.g., $v_0 \to v_1 \to v_2 \to v_3$), a high-cardinality variable may appear early in the sequence (e.g., a \texttt{rhost} at $v_1$), forcing the subtree to branch at the top layers.
Consequently, even if later variables form a highly frequent pattern (e.g., \texttt{uid=509} $\to$ \texttt{euid=0}), it is fragmented into multiple separate branches, preventing efficient compression.
\textit{Second, we require a robust mechanism to distinguish meaningful patterns from random noise.} Since variables are high-entropy, assigning a unique ID to every combination would result in a prohibitively large dictionary and compromise the compression benefits. Therefore, we must selectively encode only frequent, co-occurring variable combinations (the ``signal'') into the URT, while actively filtering out rare or random values (the ``noise'') to be handled as residuals (Sec.~\ref{sec:residual_data_processing}).

\subsubsection{Variable Reordering for Efficient Subtree Construction}
\label{sec:filtering_reordering}


To mitigate the secondary branch explosion, \alias restructures the variable subtree topology via a two-step optimization process. This involves a stability-based sort that prioritizes low-entropy variables for insertion. By forcing frequent patterns to share a common prefix, this strategy effectively pushes high-cardinality branching to the deeper levels of the tree. Consequently, correlated variable sequences are materialized as deep, continuous paths (Fig.~\ref{fig:final_structural_tree}) rather than fragmenting prematurely.

\begin{figure*}[t]
    \centering
    \includegraphics[width=1\textwidth]{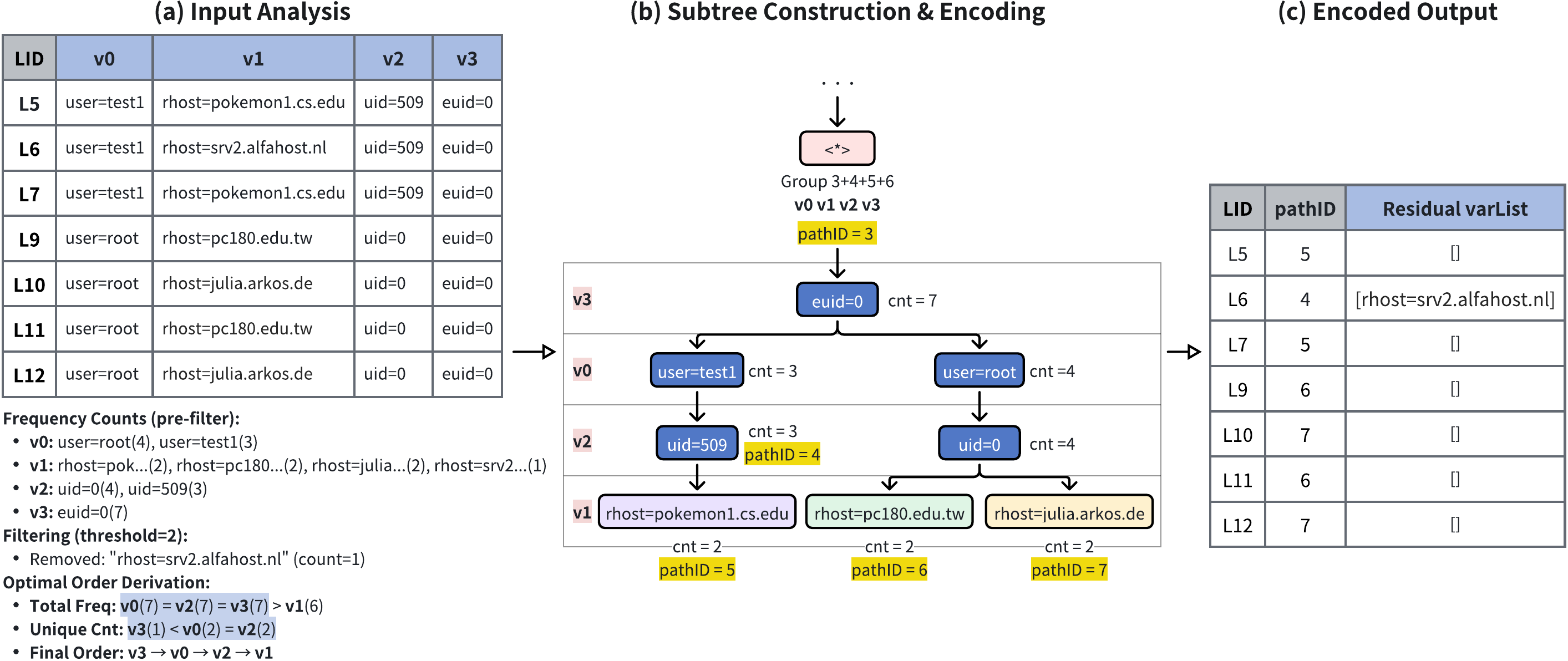}
    \caption{The Variable Subtree Encoding Pipeline for the \texttt{sshd} Log Group (\texttt{pathID=3})}
    \label{fig:variable_subtree_encoding}
\end{figure*}

The optimization begins with \textit{high-frequency filtering}, which prunes infrequent variables, as shown in Fig.~\ref{fig:variable_subtree_encoding}(a). 
For each variable position within a log group, we compute the frequency of every unique value and discard any that fall below a pre-defined threshold $\tau$, e.g., the rare \texttt{rhost=srv2.alfahost.nl}.
If all values at a given position are filtered out, the entire position is excluded from subsequent analysis.
For instance, in the \texttt{httpd} log group (\texttt{pathID=1}) shown in Fig.~\ref{fig:final_structural_tree}, the sole variable $v_0$ (\texttt{transaction\_id}) is removed entirely because both of its values (\texttt{5001} and \texttt{5002}) are too infrequent.
In this work, we use $\tau$ as our universal threshold for different filtering operations to screen out outliers, which avoids the overhead of managing multiple distinct thresholds and reduces overall system complexity.

The second step is \textit{variable reordering}, which establishes the optimal insertion order for the remaining variables.
The reordering is determined by a heuristic that prioritizes stability. We calculate two key metrics for each variable position, i.e., \textit{total frequency} (the sum of occurrences of all its values) and \textit{discriminative power} (the number of its unique values).
Variables are then sorted in descending order of total frequency. Any ties are resolved by sorting in ascending order of discriminative power, which favors variables with fewer unique values. In the \texttt{sshd} example, after filtering, $v_1$ has the lowest total frequency, i.e., six, while the other positions share a total frequency of seven. Among those, $v_3$ has the lowest discriminative power of one. This heuristic thus yields the optimal construction order of $v_3 \to v_0 \to v_2 \to v_1$, which is the crucial input for the final subtree construction phase.

\subsubsection{Subtree Construction and Unified Redundancy Encoding}

With the established variable order, this phase constructs the variable subtrees to identify and encode co-occurrence patterns. While the reordering was guided by aggregate statistics, it does not guarantee that individually frequent variables appear together in the same log entry.
For instance, a specific \texttt{user} and \texttt{rhost} might both be globally frequent, but if they never co-occur (e.g., that user never logs in from that host), they do not form a valid pattern.
Therefore, this stage validates these patterns by iterating through each log's \texttt{varList}, ensuring that compression identifiers are assigned only to variable combinations that genuinely exist in the data.

The construction process is illustrated in Fig.~\ref{fig:variable_subtree_encoding}(b).
As each log's \texttt{varList} is traversed according to the optimal order (i.e., $v_3 \to v_0 \to v_2 \to v_1$), a path is extended in the subtree. Every node in this tree maintains a \texttt{cnt} attribute, a key metric that records the number of logs whose reordered variables share that specific prefix.
For logs like L5 and L7, which are composed entirely of frequent variables, a complete path is formed, and the \texttt{cnt} value of every node along that path is incremented.
In contrast, the traversal for log L6 terminates prematurely because its variable \texttt{rhost=srv2.alfahost.nl} was filtered out as a low-frequency value. Consequently, L6 only increments the cnt values for its matched prefix (\texttt{euid=0} $\to$ \texttt{user=test1} $\to$ \texttt{uid=509}). This distinction is crucial, as the \texttt{cnt} attribute now precisely tracks the frequency of both complete and partial patterns.
Next, a pruning operation is performed, which acts as a second layer of filtering, removing paths composed of individually frequent variables but whose combination is rare. 
Any node whose $\texttt{cnt}$ falls below the pruning threshold $\tau$ is removed.
While no nodes are pruned in our simplified example (Fig.~\ref{fig:variable_subtree_encoding}), this is crucial for eliminating noise in complex datasets.

The most critical step is the assignment of new universal \texttt{pathID}s.
To maximize compression efficiency, identifiers are assigned only to nodes that represent the termination of a high-frequency aggregate pattern.
This ensures that every \texttt{pathID} corresponds to a meaningful, recurring token combination, preventing an explosion in dictionary size due to rare partial matches. We term these nodes ``stable endpoints'' and identify them using the \texttt{cnt} attribute under two conditions.
First, any leaf node is inherently a stable endpoint, as it represents the explicit termination of a pattern that has already survived the high-frequency filtering process (i.e., $\tau$).
For example, the \texttt{rhost=pokemon1.cs.edu} node in Fig.~\ref{fig:variable_subtree_encoding}(b) is assigned \texttt{pathID=5} because it marks the end of the pattern for logs L5 and L7.
Second, a non-leaf node is designated a stable endpoint if the difference between its \texttt{cnt} and the sum of its children's \texttt{cnt} values is not less than the threshold $\tau$. This ``residual count'' represents the number of log entries whose pattern matches exactly up to this node but does not continue to any of its high-frequency children. By enforcing the threshold, we ensure that we only create a new ID if a substantial number of logs terminate at this specific intermediate point. For instance, the \texttt{uid=509} node has a \texttt{cnt} of 3, while its only child has a \texttt{cnt} of 2. The residual count of 1 indicates that one log (L6) terminates its match here. If this residual meets the threshold, the node is marked as a stable endpoint and assigned \texttt{pathID=4}, allowing \alias to hierarchically encode both complete patterns and frequent sub-patterns.


The final encoded output, shown in Fig.~\ref{fig:variable_subtree_encoding}(c), is a highly compact representation of the log data. Log messages that fully match a frequent path, such as L7 and L9, are represented by a single \texttt{pathID} (5 and 6, respectively). Logs that only partially match, like L6, are represented by the \texttt{pathID} of their longest matched prefix (\texttt{pathID=4}), while the unmatched token (\texttt{rhost=srv2.alfahost.nl}) is preserved as a residual variable in the log's \texttt{varList} for processing in the third stage.
This process is the core manifestation of our unified redundancy encoding concept.
Crucially, the new \texttt{pathID} is a logical extension of the initial structural \texttt{pathID}. It represents the complete ``structure + variable'' collective pattern. Unlike traditional methods that require one identifier for the template and separate ones for each variable, our approach uses a single \texttt{pathID} to represent the entire high-frequency combination. For fully matched logs, this identifier replaces both the original structure and all of its variables, achieving a significant gain in compression efficiency.

\subsection{Residual Data Processing}
\label{sec:residual_data_processing}

After the first two stages have captured high-frequency collective patterns, the remaining data constitutes the ``long-tail'' data, i.e., outlier variables characterized by high entropy and weak correlation. This final stage is designed to efficiently compress these residual components by leveraging simpler, linear patterns that may exist. Our approach is guided by the principle of efficient residual handling: after the URT has filtered out complex correlations, this final pipeline can focus on simpler sequential regularities.



\subsubsection{Global Sorting for Temporal Coherence}

The structural grouping in the first stage, while effective for mining template-based redundancy, inevitably fragments the natural time order of logs. This can obscure valuable patterns that span across different structural groups. As shown in Fig.~\ref{fig:residual_data_processing}(a), logs from different sources (e.g., \texttt{httpd} and \texttt{database}) may share a sequential \texttt{transaction\_id}. To uncover these dependencies, the core of this stage is a \textit{global sorting pipeline}. We aggregate the residual information from all log groups and perform a global re-sort based on the original Line ID (\texttt{Lid}). This operation restores the dataset's temporal coherence, realigning dispersed entries like L3, L4, and L8 and exposing the incremental numeric sequence (\texttt{5001}, \texttt{5001}, \texttt{5002}), which is now highly amenable to delta compression.

\subsubsection{Just-in-Time Residual Templatization}

The residual variables after the second stage are a heterogeneous mix of complex strings, atomic identifiers, and pure numbers. This diversity prevents efficient columnar compression. To resolve this, we employ a \textit{just-in-time templatization} strategy that iterates through the sorted queue and parses each variable into homogeneous components.

For each residual variable, a regex-based extraction mechanism decomposes it into its invariant static fragments and dynamic numeric parts. We then borrow the core principle from Denum~\cite{DBLP:conf/kbse/YuW0H24} to classify each extracted numeric string based on its intrinsic features (e.g., length, first digit) and generate a corresponding placeholder. This unified process gracefully handles three distinct scenarios:

\begin{itemize}[noitemsep,leftmargin=5.5mm]
    \item \textbf{Complex Numeric Variables}: For a composite variable like \texttt{audit(1119799950.864:693295):} in log L1, the process generates a generalized template from its static parts and placeholders (e.g., \texttt{audit(}$\langle jb\_ \rangle$\texttt{.}$\langle c\_ \rangle$\texttt{:}$\langle fg\_ \rangle$\texttt{):}) and dispatches each numeric value to its respective columnar stream.

    \item \textbf{Atomic String Variables}: For a variable that cannot be further deconstructed, such as \texttt{rhost=srv2.} \texttt{alfahost.nl} from log L6, the entire string is treated as an indivisible atomic template and assigned a unique template ID.

    \item \textbf{Pure Numeric Variables}: For simple numbers like \texttt{5001}, a minimalist placeholder template (e.g., $\langle df\_ \rangle$) is generated, and the value is appended to the corresponding numeric stream.
\end{itemize}

\begin{figure*}[t]
    \centering
    \includegraphics[width=0.8\textwidth]{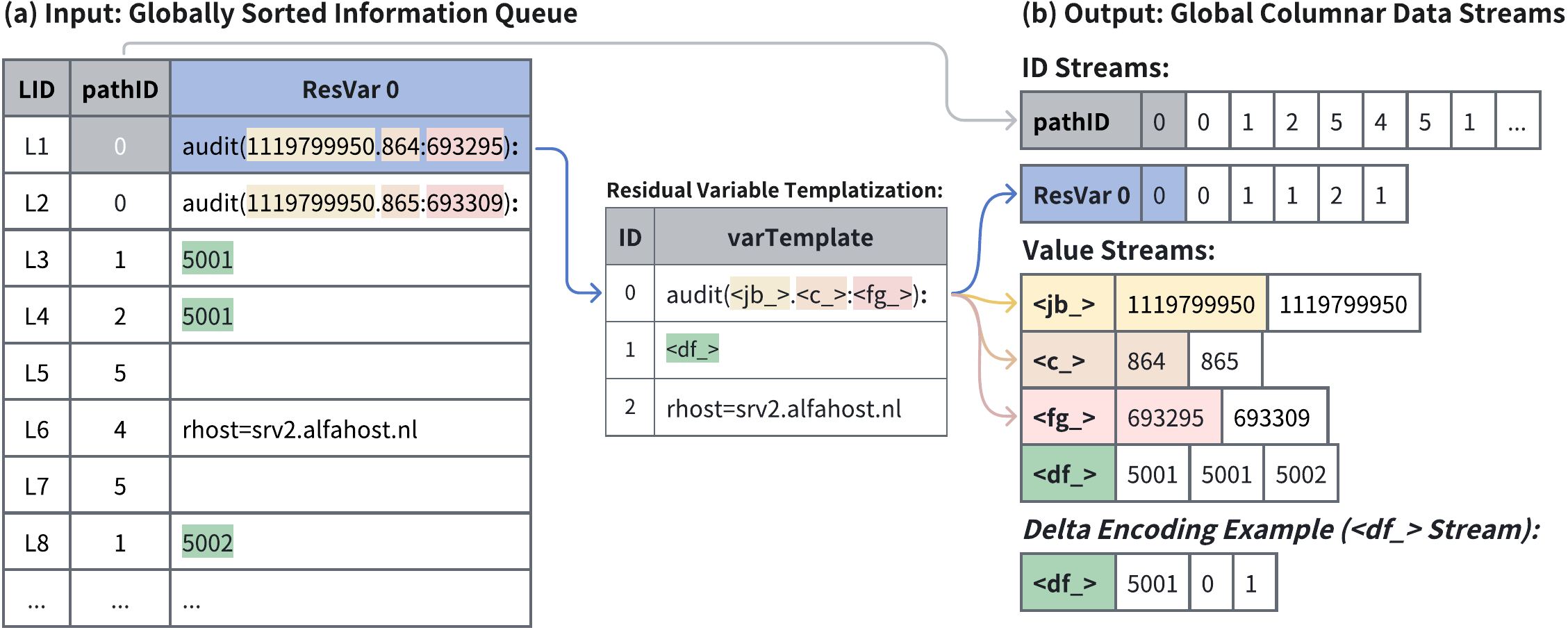}
    \caption{The Residual Data Processing Pipeline in Stage 3}
    \label{fig:residual_data_processing}
\end{figure*}

The result of this process is a set of dense, homogeneous columnar streams, as shown in Fig.~\ref{fig:residual_data_processing}(b). The \texttt{pathID} of each log is written to the main stream, followed by the template IDs for its residual variables in the \texttt{ResVar} streams. Concurrently, all extracted numeric values are written to their specialized Value Streams, which are then compressed using Delta Encoding followed by ZigZag and Varint encoding.

Although this stage borrows the principle of numeric classification from Denum, our hierarchical redundancy mining strategy is fundamentally different. Denum performs an indiscriminate, global numeric/non-numeric split from the outset, which can prematurely destroy valuable contextual correlations. In contrast, \alias has already processed the vast majority of high-frequency tokens (including many numerics) via unified redundancy encoding
in the first two stages. Consequently, the computationally intensive parsing logic of this stage is applied only to a minimal subset of true residual data. This targeted approach provides a dual advantage: a higher compression ratio, as it preserves the ``structure + variable'' correlations, and faster compression speed, by confining the expensive variable processing to a much smaller dataset.

The pipeline concludes by aggregating all generated components, i.e., the URT, the residual template dictionary, and the various columnar data streams, into a single compact archive, which is finally processed by a general-purpose compressor (e.g., LZMA) to exploit remaining byte-level redundancy.

\subsection{Decompressor}

The decompression process of \alias is the precise inverse of compression, achieving lossless log reconstruction. The process begins by loading the metadata from the compressed archive, including the URT, the residual template dictionary, and all columnar data streams. The core of the reconstruction is a sequential traversal of the main \texttt{pathID} stream, reconstructing logs line by line. Since this stream was written in the order of the original log entries, processing it sequentially naturally restores the original log order.

For each \texttt{pathID} read from the stream, a two-stage reconstruction process is performed. First, the \texttt{pathID} is used to perform a reverse lookup in the URT. By tracing backward from the \texttt{pathID} node to the root, we instantly reconstruct the log's complete high-frequency ``structure + variable'' pattern. For a log that was fully matched during compression (e.g., L5), this single lookup is sufficient to recover all its tokens, including both the structural and variable tokens.
For a log with residual variables (e.g., L1 or L6), this lookup only recovers the matched portion, and the remaining variables are then reconstructed by reading their template IDs from the \texttt{ResVar} streams and resolving them against the residual template dictionary.
At this point, a semi-reconstructed log line is formed, containing all static and high-frequency variable tokens but still holding placeholders for dynamic values (e.g., $\langle dt \rangle$, $\langle P \rangle$, and $\langle jb\_ \rangle$). Therefore, the second step performs global placeholder substitution. For each placeholder encountered, the decompressor reads the next available value from the corresponding columnar data stream and substitutes it in place. This multi-layered mechanism guarantees 100\% accuracy, as all data streams are consumed in the exact order in which they were generated, ensuring a perfect reconstruction of the original log file.

\section{Evaluation}
\label{sec:evaluation}

We conduct a comprehensive evaluation to demonstrate the effectiveness and efficiency of \alias. Our evaluation is designed to answer four critical research questions:

\begin{itemize}[noitemsep,leftmargin=5.5mm]
    \item \textbf{RQ1:} How does \alias's compression ratio compare to state-of-the-art log compressors?
    \item \textbf{RQ2:} How does \alias's compression speed compare to state-of-the-art log compressors?
    \item \textbf{RQ3:} What are the individual performance contributions of \alias's hierarchical stages?
    \item \textbf{RQ4:} What is the impact of input data granularity on \alias's performance?
\end{itemize}

\subsection{Experimental Settings}
\label{sec:experimental_setting}

\subsubsection{Datasets}

Consistent with our previous empirical study (Sec.\ref{sec:empirical_study}), we utilize the LogHub benchmark~\cite{DBLP:conf/icse/ZhuHLHXZL19} for evaluation. However, while the empirical study involved only 14 datasets, this evaluation incorporates the full suite of 16 datasets. The additional datasets (i.e., Android and Windows) were previously excluded due to the scalability limitations of certain parsers. In this evaluation phase, all selected compressors successfully handle these datasets using their default parsers, allowing for a comprehensive assessment across the entire benchmark.

\subsubsection{Evaluation Metrics}

We employ two fundamental metrics in the field of log compression: the \textbf{Compression Ratio (CR)} to quantify effectiveness (higher values indicate better compression) and the \textbf{Compression Speed (CS)} to measure efficiency (higher values indicate faster processing). The CR was previously defined in Sec.~\ref{sec:background}. The CS is defined as:

\begin{equation*}
    Compression~Speed~(CS) = \frac{Original~File~Size~(MB)}{Compression~Time~(s)}
\end{equation*}

\subsubsection{Baselines}

We compare \alias against comprehensive baselines spanning both state-of-the-art log-specific compressors and well-known general-purpose compressors.



\begin{itemize}[noitemsep,leftmargin=5.5mm]
    \item \textbf{Log-Specific Compressors:} We evaluate \alias against the same four state-of-the-art methods studied in our empirical study (Sec.~\ref{sec:empirical_study}). These include LogZip~\cite{DBLP:conf/kbse/LiuZHHZL19}, the pioneering parser-based method; LogReducer~\cite{DBLP:conf/fast/WeiZWLZCSZ21} and LogShrink~\cite{DBLP:conf/icse/LiZL024}, which represent advanced approaches optimized for parameter correlation and variability; and Denum~\cite{DBLP:conf/kbse/YuW0H24}, the leading number-centric compressor that operates without traditional parsing. This selection ensures a comprehensive comparison against the best existing techniques across different design paradigms.
    

    \item \textbf{General-Purpose Compressors:} We include four compressors in this category: \textit{gzip} (a traditional compressor based on the DEFLATE algorithm, known for good speed but moderate compression ratios), \textit{bzip2} (which uses the Burrows-Wheeler Transform for better compression at the cost of lower speed), \textit{LZMA} (a dictionary-based algorithm typically offering the best compression ratios but is relatively slow), and \textit{PPMd} (which compresses by analyzing character sequences and predicting probabilities, excelling on plain text but degrading significantly on logs with high-entropy numeric variables).
\end{itemize}

For metrics independent of the experimental environment (CR), we directly cite the results from the original papers of the respective log compressors. For environment-dependent metrics (CS), we re-ran all available open-source tools in our unified environment to ensure fair comparison.

\subsubsection{Implementation}

All experiments were conducted on a Linux server with an AMD EPYC 9224 CPU (24 cores/48 threads, 3.70 GHz) and 251 GB RAM, running Ubuntu 22.04 LTS (kernel 6.8.0). \alias is implemented in C++ and compiled with g++ following the C++20 standard. Regular expression matching relies on the PCRE2 library. Similar to Denum, the \texttt{PCRE2\_CODE\_UNIT\_WIDTH} was set to 8.

To align with the evaluation settings of Denum and related work, all input log data for RQ1-RQ3 experiments were split into 100K-line chunks for parallel compression. 
Each compressor uses 4 threads for processing each chunk. 
To ensure fair architectural comparison, our main evaluation of \textbf{\alias} constrains the internal operations of each chunk-processing thread to be single-threaded, matching the execution model of the baselines. To demonstrate the full potential of our parallel-aware design, we also evaluate an enhanced configuration, denoted as \textbf{\alias-P}, where each of the 4 chunk-processing threads utilizes an internal pool of 4 worker threads.
The total time to compress all chunks is recorded.
\alias employs the `tar' utility to package all generated intermediate files, which are then compressed with `lzma' to form the final archive.

\subsection{RQ1: The Compression Ratio of \alias}
\label{sec:rq1_cr}

Table~\ref{tab:rq1_cr} presents the compression ratio comparison across all benchmarks. \alias achieves the highest compression ratio on 14 of 16 datasets, establishing a new state-of-the-art in compression effectiveness.



\noindent \textbf{Comparison with general-purpose compressors:}
Regarding general-purpose tools, \alias demonstrates substantial superiority over all of them. Compared to gzip, \alias achieves a 5.68$\times$ higher average compression ratio, reaching up to 27.94$\times$ on specific datasets.
Against the stronger LZMA baseline, improvements range from 1.57$\times$ (Proxifier) to 5.99$\times$ (OpenSSH).
Similarly, the ratio over bzip2 ranges from 1.26$\times$ (Proxifier) to 7.36$\times$ (Windows), and over PPMd, it ranges from 1.17$\times$ (Proxifier) to 8.14$\times$ (Windows).
These results confirm that exploiting log-specific structure yields significant compression gains.


\noindent \textbf{Comparison with log-specific compressors:} 
\alias advances the state-of-the-art among log-specific methods. Compared to LogReducer, CR improvements reach up to 59.26\%. Against LogShrink, the maximum gain is 56.63\%. Notably, even compared to the current leading method Denum, \alias achieves substantial improvements of up to 25.93\% on Linux and 18.86\% on Thunderbird. The LogZip result for Thunderbird is marked as unavailable (--) because, as reported in the LogShrink paper~\cite{DBLP:conf/icse/LiZL024}, LogZip failed to complete parsing within one week.


\begin{table*}[t]
    \centering
    \small
    \caption{Experimental Results of Compression Ratio}
    \label{tab:rq1_cr}
    \begin{tabular}{l c c c c c c c c c}
        \toprule
        \textbf{Dataset} & \textbf{gzip} & \textbf{LZMA} & \textbf{bzip2} & \textbf{PPMd} & \textbf{LogZip} & \textbf{LogReducer} & \textbf{LogShrink} & \textbf{Denum} & \textbf{\alias} \\
        \midrule
        \midrule
        Android & 7.742 & 18.857 & 12.787 & 19.370 & 25.165 & 20.776 & 21.857 & 32.494 & \textbf{32.822} \\
        Apache & 21.308 & 25.186 & 29.557 & 31.688 & 30.375 & 43.028 & 55.940 & 58.517 & \textbf{64.330} \\
        BGL & 12.927 & 17.637 & 15.461 & 18.927 & 32.655 & 38.600 & 42.385 & 41.804 & \textbf{47.145} \\
        Hadoop & 20.485 & 36.095 & 32.598 & 32.110 & 35.008 & 52.830 & 60.091 & 78.546 & \textbf{79.733} \\
        HDFS & 10.636 & 13.559 & 14.059 & 19.155 & 26.666 & 22.634 & \textbf{27.319} & 25.670 & 26.466 \\
        HealthApp & 10.957 & 13.431 & 13.843 & 15.337 & 12.632 & 31.694 & 39.072 & 44.472 & \textbf{50.477} \\
        HPC & 11.263 & 15.076 & 12.756 & 14.822 & 27.208 & 32.070 & 35.878 & 45.275 & \textbf{45.447} \\
        Linux & 11.232 & 16.677 & 14.695 & 18.508 & 23.368 & 25.213 & 29.252 & 30.449 & \textbf{38.343} \\
        Mac & 11.733 & 22.159 & 18.074 & 28.469 & 26.306 & 35.251 & 39.860 & 40.789 & \textbf{44.992} \\
        OpenSSH & 16.828 & 18.918 & 22.865 & 31.977 & 42.606 & 86.699 & 103.175 & 101.654 & \textbf{113.327} \\
        OpenStack & 12.158 & 14.437 & 15.231 & 17.429 & 17.258 & 16.701 & 22.157 & 22.238 & \textbf{23.356} \\
        Proxifier & 15.716 & 18.982 & 23.619 & 25.489 & 21.493 & 25.501 & 27.029 & 27.288 & \textbf{29.849} \\
        Spark & 17.825 & 19.908 & 26.497 & 30.614 & 20.825 & 59.470 & \textbf{59.739} & 59.470 & 54.813 \\
        Thunderbird & 16.462 & 27.309 & 25.428 & 33.026 & -- & 49.185 & 48.434 & 63.824 & \textbf{75.862} \\
        Windows & 17.798 & 202.568 & 67.533 & 61.083 & 310.596 & 342.975 & 456.301 & 481.350 & \textbf{497.298} \\
        Zookeeper & 25.979 & 27.667 & 36.156 & 38.931 & 47.373 & 94.562 & 116.981 & 135.251 & \textbf{143.742} \\
        \bottomrule
    \end{tabular}
\end{table*}

This systematic performance advantage of \alias stems from a fundamental architectural shift.
Traditional methods (whether the ``parse-then-compress'' workflows or Denum's ``numeric/non-numeric'' split) perform an irreversible token categorization early in the process.
This leads to the loss of contextual correlations that spans these category boundaries.
In contrast, \alias's unified redundancy encoding paradigm co-designs structural extraction with pattern encoding.
This enables representing entire, highly correlated ``structure + variable'' token sequences with single path identifiers, where other methods require one template ID plus multiple independent variable IDs.
Our hierarchical redundancy mining strategy prioritizes encoding high-value aggregate patterns while preserving full context, allowing it to exploit deep redundancies in log data.

On the two datasets where \alias does not achieve the highest CR (HDFS and Spark), performance remains highly competitive. On HDFS, the difference from optimal is minimal (3.12\%). The larger gap on Spark reflects the characteristics of the dataset, i.e., Spark logs are dominated by simple, repetitive templates with single continuously changing numeric values (e.g., over a million records like \texttt{"Update row <*>"})~\cite{DBLP:conf/kbse/YuW0H24}. In this case, the deep correlation mining of \alias's second stage offers less advantage compared to methods specifically optimized for simple numeric streams. Nevertheless, \alias achieves a better performance balance in these cases with significantly higher compression speed, as we demonstrate in RQ2.


\begin{summarybox}
\textbf{Summary for RQ1:}
    \alias surpasses existing general-purpose and log-specific compressors in compression ratio on 14 out of 16 datasets, establishing a new state-of-the-art in compression effectiveness.
\end{summarybox}

\subsection{RQ2: The Compression Speed of \alias}
\label{sec:rq2_cs}

We evaluate \alias's efficiency in two configurations: \textbf{\alias}, which restricts internal processing to a single thread per chunk for a fair comparison, and \textbf{\alias-P}, which activates our internal fine-grained parallelism.
A critical aspect of this benchmark is the timing methodology. We measure the processing time for \alias (both configurations) and Denum on a strictly end-to-end basis. In contrast, for parser-based baselines (LogZip, LogReducer, LogShrink), we adhere to the conventions established in their respective papers, which exclude the often time-consuming parsing and template generation phases. Consequently, \alias's performance is achieved under a significantly stricter measurement standard than the parser-based baselines.


Table~\ref{tab:rq2_cs} presents the experimental results, which reveal two key conclusions.
First, in the direct comparison (single-threaded internal model), \alias emerges as the fastest compressor, outperforming all baselines across all 16 datasets.
With an average speed of 29.87 MB/s, it surpasses the next fastest competitor, Denum (17.83 MB/s), by 1.68$\times$.
The superior efficiency stems from \alias's hierarchical processing strategy. By efficiently discovering and encoding frequent ``structure + variable'' patterns in the second stage, we drastically reduce the volume of data requiring processing by the third stage that is more computationally intensive.
Second, the results for \alias-P validate the significant benefits of our parallel-aware design. By enabling internal, fine-grained parallelism, \alias-P achieves an average speed of 41.55 MB/s, representing a further 39.1\% average speedup over the single-threaded \alias configuration.

The observed variation in \alias's speed across datasets is primarily driven by dataset size, which dictates the effectiveness of chunk-level parallelization. For example, on the small Linux dataset (25,567 lines), the volume is insufficient to fill even a single 100K-line processing chunk, precluding the standard multi-threaded acceleration used by all compressors.
Despite this constraint, the standard \alias configuration achieves the best performance (8.55 MB/s) among all baselines. Moreover, \alias-P provides substantial additional speedup by leveraging the co-design of hierarchical redundancy mining and internal fine-grained parallelism. This ensures high efficiency even in edge cases. On the Linux dataset, \alias-P (10.23 MB/s) is significantly faster than Denum (4.97 MB/s) because its hierarchical strategy minimizes the computational overhead, while the internal worker threads maximize CPU utilization within the single active chunk.

\begin{table*}[t]
    \centering
    \small
    \caption{Experimental Results of Compression Speed (MB/s)}
    \label{tab:rq2_cs}
    \begin{tabular}{l c c c c c | c}
        \toprule
        \textbf{Dataset} & \textbf{LogZip} & \textbf{LogReducer} & \textbf{LogShrink} & \textbf{Denum} & \textbf{\alias} & \textbf{\alias-P} \\
        \midrule
        \textbf{Implementation} & Python & C++ & C++ and Python & C++ & C++ & C++ \\
        \midrule
        Android   & 0.068 & 19.323 & 4.123 & 24.380 & \textbf{31.189} & 36.672 \\
        Apache    & 0.737 & 2.347  & 1.537 & 6.369  & \textbf{12.002} & 17.551 \\
        BGL       & 0.874 & 26.738 & 2.571 & 23.575 & \textbf{33.234} & 50.625 \\
        Hadoop    & 0.901 & 12.882 & 4.401 & 24.453 & \textbf{40.539} & 52.840 \\
        HDFS      & 0.701 & 23.598 & 3.466 & 25.193 & \textbf{29.281} & 35.288 \\
        HealthApp & 0.736 & 7.937  & 2.754 & 17.866 & \textbf{22.239} & 29.103 \\
        HPC       & 0.644 & 9.391  & 3.599 & 25.216 & \textbf{32.257} & 37.868 \\
        Linux     & 0.687 & 1.249  & 0.941 & 4.969  & \textbf{8.550}  & 10.228 \\
        Mac       & 0.009 & 5.450  & 2.141 & 6.710  & \textbf{11.715} & 16.097 \\
        OpenSSH   & 0.715 & 14.773 & 3.335 & 25.572 & \textbf{51.369} & 70.016 \\
        OpenStack & 0.537 & 13.039 & 4.018 & 12.352 & \textbf{20.960} & 29.396 \\
        Proxifier & 0.716 & 1.328  & 0.742 & 6.000  & \textbf{9.657}  & 10.585 \\
        Spark     & 0.550 & 21.233 & 3.185 & 26.034 & \textbf{42.052} & 59.005 \\
        Thunderbird & --  & 18.656 & 4.036 & 19.830 & \textbf{38.539} & 69.213 \\
        Windows   & 1.357 & 18.483 & 6.330 & 28.783 & \textbf{79.220} & 116.152 \\
        Zookeeper & 0.842 & 4.429  & 2.280 & 8.000  & \textbf{15.111} & 24.192 \\
        \midrule
        \textbf{Average} & 0.694 & 12.554 & 3.091 & 17.831 & \textbf{29.870} & 41.552 \\
        \bottomrule
    \end{tabular}
\end{table*}

\begin{summarybox}
\textbf{Summary for RQ2:}
    \alias achieves state-of-the-art end-to-end compression speed. This efficiency stems from the synergistic effect of its hierarchical processing strategy, which minimizes computational overhead, and its parallel-aware architecture, which maximizes resource utilization.
\end{summarybox}

\subsection{RQ3: Ablation Analysis of \alias's Different Stages}
\label{sec:rq3_ablation}


We perform a progressive ablation study to isolate and quantify the contribution of each stage in \alias's hierarchical design.
Specifically, we focus on Stage 2 (Variable Subtree Encoding), which implements our core unified redundancy encoding paradigm. We evaluate three configurations across all 16 datasets: (1) LZMA, a general-purpose compressor serving as a universal baseline; (2) \alias (S1+S3), a structural baseline that executes Stage 1 and Stage 3 but bypasses Stage 2, representing an advanced parser-based compressor that treats structure and variables as separate entities; and (3) \alias, the complete model, which is compared against \alias (S1+S3) to measure the specific performance gain provided by mining ``structure + variable'' correlations.
Both the last two configurations utilize LZMA for the final compression of their output.
The results for CR and CS are detailed in Tables~\ref{tab:rq3_cr} and~\ref{tab:rq3_cs}, respectively.

\noindent \textbf{Compression Ratio Analysis:}
The results demonstrate a clear, step-wise improvement in effectiveness.
The \alias (S1+S3) model is itself a high-performance compressor. By leveraging the structural analysis of Stage 1 and global sorting pipeline of Stage 3, it achieves a 2.48$\times$ higher average CR than LZMA. Furthermore, the integration of Stage 2 delivers a decisive performance boost. The full \alias model achieves an additional $8.28\%$ increase in average CR over the already strong \alias (S1+S3) variant. This gain is most prominent on datasets with deep variable correlations, such as Android (+28.39\%) and Thunderbird (+26.70\%).
Crucially, the full model consistently outperforms the variant baseline across all evaluated datasets, ensuring robust improvements regardless of log characteristics. This validates our unified redundancy encoding design, i.e., by treating variable combinations as part of a compressible pattern rather than independent values, \alias unlocks massive redundancy in log data.

\begin{table}[t]
    \centering
    \small
    \caption{Ablation Study of Compression Ratio}
    \label{tab:rq3_cr}
    \begin{tabular}{l c c c}
        \toprule
        \textbf{Dataset} & \textbf{LZMA} & \textbf{\alias (S1+S3)} & \textbf{\alias} \\
        \midrule
        \midrule
        Android & 18.857 & 25.564 & \textbf{32.822} \\
        Apache & 25.186 & 61.887 & \textbf{64.330} \\
        BGL & 17.637 & 46.548 & \textbf{47.145} \\
        Hadoop & 36.095 & 75.170 & \textbf{79.733} \\
        HDFS & 13.559 & 24.282 & \textbf{26.466} \\
        HealthApp & 13.431 & 46.932 & \textbf{50.477} \\
        HPC & 15.076 & 43.880 & \textbf{45.447} \\
        Linux & 16.677 & 37.356 & \textbf{38.343} \\
        Mac & 22.159 & 37.373 & \textbf{44.992} \\
        OpenSSH & 18.918 & 95.969 & \textbf{113.327} \\
        OpenStack & 14.437 & 21.868 & \textbf{23.356} \\
        Proxifier & 18.982 & 26.664 & \textbf{29.849} \\
        Spark & 19.908 & 50.656 & \textbf{54.813} \\
        Thunderbird & 27.309 & 59.876 & \textbf{75.862} \\
        Windows & 202.568 & 490.787 & \textbf{497.298} \\
        Zookeeper & 27.667 & 118.531 & \textbf{143.742} \\
        \midrule
    \end{tabular}
\end{table}

\noindent \textbf{Compression Speed Analysis:}
The ablation study also reveals the critical role of Stage 2 as a performance accelerator. Despite adding another analysis step, the full \alias model is $25.93\%$ faster on average than the structurally simpler \alias (S1+S3) variant. This counter-intuitive result validates our hierarchical redundancy mining strategy. In the \alias (S1+S3) configuration, every variable identified in Stage 1 must be processed by Stage 3, which relies on computationally expensive regex templatization.
In contrast, the full \alias model only supplies Stage 3 with a smaller set of ``true long-tail residuals.''
By handling the majority of high-frequency variables in the efficient Stage 2, \alias drastically reduces the workload of the most time-consuming part of the pipeline, thereby lowering overall computational overhead and increasing throughput.


\begin{table}[t]
    \centering
    \small
    \caption{Ablation Study of Compression Speed (MB/s)}
    \label{tab:rq3_cs}
    \begin{tabular}{l c c c}
        \toprule
        \textbf{Dataset} & \textbf{LZMA} & \textbf{\alias (S1+S3)} & \textbf{\alias} \\
        
        \midrule
        \midrule
        
        Android & 24.876 & 26.717 & \textbf{36.672} \\
        Apache & 7.665 & 13.453 & \textbf{17.551} \\
        BGL & 16.108 & 49.072 & \textbf{50.625} \\
        Hadoop & 22.102 & 44.937 & \textbf{52.840} \\
        HDFS & 15.715 & 31.145 & \textbf{35.288} \\
        HealthApp & 9.436 & 23.156 & \textbf{29.103} \\
        HPC & 11.700 & 33.401 & \textbf{37.868} \\
        Linux & 4.592 & 8.818 & \textbf{10.228} \\
        Mac & 8.428 & 12.950 & \textbf{16.097} \\
        OpenSSH & 15.873 & 46.553 & \textbf{70.016} \\
        OpenStack & 11.647 & 20.578 & \textbf{29.396} \\
        Proxifier & 6.696 & 8.750 & \textbf{10.585} \\
        Spark & 20.757 & 50.734 & \textbf{59.005} \\
        Thunderbird & 26.845 & 51.360 & \textbf{69.213} \\
        Windows & 63.706 & 91.402 & \textbf{116.152} \\
        Zookeeper & 6.581 & 14.907 & \textbf{24.192} \\
        \midrule
        \textbf{Average} & 17.045 & 32.996 & \textbf{41.552} \\
        \bottomrule
    \end{tabular}
\end{table}

\begin{summarybox}
\textbf{Summary for RQ3:}
    The ablation study confirms that \alias's performance breakthrough is driven by its hierarchical design. While Stages 1 and 3 provide a solid structural baseline, the core innovation (i.e., Stage 2 Variable Subtree Encoding) delivers a decisive performance boost to both compression ratio and speed.
\end{summarybox}

\subsection{RQ4: Robustness Analysis: The Impact of Data Granularity}
\label{sec:rq4_robustness}

This question investigate the impact of data granularity on \alias's performance, which is a critical operational parameter. Our primary experiments utilized a fixed chunk size of $100\text{K}$ lines, a standard baseline setting that maximizes parallelism but limits pattern discovery to local windows. This design involves an inherent trade-off: each chunk builds an independent URT, preventing the sharing of patterns across the full dataset.
This can potentially obscure very long-range redundancies. To explore this speed-vs-compression trade-off, we configure \alias to operate in single-archive (non-chunked) mode, allowing it to construct a single global URT over the entire dataset. We compare this configuration against Denum, the leading state-of-the-art baseline that also supports global operation.
The results, presented in Table~\ref{tab:rq4_granularity}, demonstrate that global pattern discovery significantly boosts \alias's compression effectiveness while maintaining highly competitive speeds.

\noindent \textbf{Compression Ratio Analysis:}
In single-archive mode, \alias achieves a 273.27\% increase in average CR over its own chunked performance, with the gain on the Windows dataset reaching a massive 739.37\%.
When compared to Denum operating in the same global configuration,
\alias outperforms it on 13 of the 16 datasets. The improvements are particularly significant on complex logs such as Windows (+22.49\%) and Thunderbird (+21.29\%). On two of the remaining three datasets (Hadoop, HDFS), \alias remains highly competitive, following Denum with a small difference of less than 2.50\%. The only notable exception is the Spark dataset, where Denum's specialized handling of specific numeric sequences proves more efficient. Overall, these results validate that \alias scales effectively to leverage global context to achieve superior compression.

\noindent \textbf{Compression Speed Analysis:}
\alias maintains an absolute speed advantage even when building a single global model, whose average speed of 15.77 MB/s is 2.62$\times$ faster than Denum (6.01 MB/s). This sustained high performance is attributable to \alias's hybrid parallel architecture.
While standard chunk-based compressors (like Denum) lose parallelism when processing a single global block, \alias retains fine-grained concurrency: the pre-processing of logs in Stage 1 and the subtree construction for distinct structural groups in Stage 2 continue to execute in parallel. Furthermore, the architectural benefit identified in RQ3, where Stage 2 acts as a high-throughput filter to reduce the workload of Stage 3, remains fully effective. Particularly, even in this globally optimized mode, \alias remains significantly faster than the default chunked configurations of LogReducer (12.55 MB/s) and LogShrink (3.09 MB/s).

\begin{table}[t]
    \centering
    \small
    \caption{Robustness Analysis of Single-Archive Mode Performance}
    \label{tab:rq4_granularity}
    \begin{tabular}{l c c c c}
        \toprule
        \textbf{Dataset} & \multicolumn{2}{c}{\textbf{Compression Ratio}} & \multicolumn{2}{c}{\textbf{Speed (MB/s)}} \\
        \cmidrule(lr){2-3} \cmidrule(lr){4-5}
        \textbf{} & \textbf{Denum} & \textbf{\alias} & \textbf{Denum} & \textbf{\alias} \\
        
        \midrule
        \midrule
        
        Android & 46.191 & \textbf{52.429} & 5.416 & \textbf{15.294} \\
        Apache & 58.562 & \textbf{64.269} & 5.810 & \textbf{12.369} \\
        BGL & 44.284 & \textbf{47.429} & 6.100 & \textbf{16.345} \\
        Hadoop & \textbf{92.899} & 90.590 & 8.643 & \textbf{27.186} \\
        HDFS & \textbf{32.919} & 32.198 & 5.230 & \textbf{7.956} \\
        HealthApp & 47.279 & \textbf{53.574} & 7.072 & \textbf{12.728} \\
        HPC & 45.542 & \textbf{46.350} & 6.369 & \textbf{16.276} \\
        Linux & 30.688 & \textbf{38.216} & 4.582 & \textbf{8.456} \\
        Mac & 43.633 & \textbf{50.266} & 5.245 & \textbf{11.171} \\
        OpenSSH & 106.984 & \textbf{120.116} & 7.426 & \textbf{27.962} \\
        OpenStack & 21.898 & \textbf{23.394} & 5.129 & \textbf{12.576} \\
        Proxifier & 27.141 & \textbf{29.739} & 5.484 & \textbf{9.894} \\
        Spark & \textbf{65.125} & 56.196 & 5.398 & \textbf{13.237} \\
        Thunderbird & 70.161 & \textbf{85.097} & 4.263 & \textbf{15.900} \\
        Windows & 3407.766 & \textbf{4174.176} & 6.421 & \textbf{26.897} \\
        Zookeeper & 135.846 & \textbf{142.290} & 7.556 & \textbf{18.013} \\
        \midrule
        \textbf{Average} & 267.307 & \textbf{319.146} & 6.009 & \textbf{15.766} \\
        \bottomrule
    \end{tabular}
\end{table}

\begin{summarybox}
\textbf{Summary for RQ4:}
    \alias's design offers both robustness and flexibility, providing users with a clear speed-vs-compression trade-off: the default chunked mode maximizes speed, while the single-archive mode maximizes the compression ratio at a reasonable efficiency cost. In both operational paradigms, \alias's performance advantages significantly surpass baseline models.
\end{summarybox}
\section{Related Work}
\label{sec:related_work}

This section discusses the landscape of compression techniques relevant to log data, organized by their foundational design philosophies.

\subsection{General-purpose Compression Approaches}
These algorithms eliminate statistical redundancy in data for compression, which can be categorized into three primary families. Dictionary-based approaches, exemplified by LZMA, achieve compression by replacing repeated byte sequences with references to a dictionary. Prediction-based methods, such as PPMd, leverage statistical models to encode characters based on their preceding context. Block-sorting algorithms, notably bzip2, utilize the Burrows-Wheeler Transform to cluster similar characters, thereby enhancing subsequent encoding efficiency.
While these methods exhibit strong performance on generic text data, they fundamentally operate on undifferentiated byte streams without awareness of log structure. Consequently, their effectiveness diminishes significantly when confronted with logs containing high-entropy variables such as unique request IDs and dynamic numerical values, as they cannot exploit the inherent semi-structured redundancy that spans log entries.

\subsection{Log-specific Compression Approaches}
The dominant methodology in log compression is the parser-based paradigm~\cite{DBLP:conf/kbse/LiuZHHZL19, DBLP:conf/fast/WeiZWLZCSZ21, DBLP:conf/icse/LiZL024, DBLP:conf/osdi/RodriguesLY21, DBLP:conf/eurosys/WeiZ00ZSWJ23}, which decouples structure extraction from variable encoding. LogZip established this framework by combining the Drain parser with iterative clustering. Subsequent works focused on optimizing the separated components: LogReducer introduced delta encoding for timestamps, while LogShrink utilized entropy-based analysis to identify variable patterns. Additionally, systems like CLP and LogGrep have extended this paradigm to enable direct search on compressed data.
To avoid the overhead of explicit template extraction, alternative approaches have emerged. Early works, such as LogArchive~\cite{DBLP:conf/sigmod/ChristensenL13} and MLC~\cite{DBLP:conf/trustcom/FengWL16}, applied similarity grouping or block-level deduplication to compress logs. More recently, the number-centric paradigm, exemplified by Denum~\cite{DBLP:conf/kbse/YuW0H24}, performs a global binary classification to split tokens into numeric and non-numeric streams. 

However, state-of-the-art methods share a fundamental limitation: the reliance on early, rigid token categorization. Parser-based methods separate templates from variables, while number-centric methods separate numerics from strings. This irreversible decoupling destroys contextual correlations that span these boundaries. Instead, \alias unifies structural extraction and pattern encoding to mine deep ``structure+variable'' redundancies that existing methods ignore.

\section{Conclusion}
\label{sec:conclusion}

This paper reevaluates the prevailing ``parse-then-compress'' paradigm in log storage, identifying the rigid decoupling of structure extraction and data encoding as a fundamental bottleneck. Our empirical analysis confirms that high parsing accuracy does not guarantee compression efficiency. Instead, it often obscures deep ``template-variable'' and inter-variable correlations essential for maximizing storage density.
To address this limitation, we introduce \alias, a framework that resolves this misalignment through Unified Redundancy Encoding. By dynamically modeling log structure and variable patterns within a Unified Redundancy Tree (URT), \alias effectively bridges the gap between static templates and dynamic parameters. Leveraging a hierarchical redundancy mining strategy and fine-grained parallelism, our approach simultaneously optimizes for compression ratio and throughput. 
Extensive evaluations on 16 benchmark datasets demonstrate that \alias establishes a new state-of-the-art. It achieves the highest compression ratio on 14 datasets, outperforming existing baselines by 6.12\%$\sim$83.34\%, while delivering the fastest processing speeds (1.68$\times$$\sim$43.04$\times$ faster than competitors). Furthermore, in single-archive mode, \alias boosts compression by 273.27\%, outperforming Denum by 19.39\% with a 2.62$\times$ speed advantage.
These findings demonstrate that co-designing parsing and compression is critical for unlocking the full potential of log data reduction in large-scale systems.

\section*{Data Availability}
The source code of \alias is publicly available on \href{https://github.com/Lycc42/LogPrism}{https://github.com/Lycc42/LogPrism}.


\bibliographystyle{ieeetr}
\bibliography{bibliography}

@inproceedings{DBLP:conf/kbse/LiuZHHZL19,
  author       = {Jinyang Liu and
                  Jieming Zhu and
                  Shilin He and
                  Pinjia He and
                  Zibin Zheng and
                  Michael R. Lyu},
  title        = {Logzip: Extracting Hidden Structures via Iterative Clustering for
                  Log Compression},
  booktitle    = {34th {IEEE/ACM} International Conference on Automated Software Engineering,
                  {ASE} 2019, San Diego, CA, USA, November 11-15, 2019},
  pages        = {863--873},
  publisher    = {{IEEE}},
  year         = {2019},
  url          = {https://doi.org/10.1109/ASE.2019.00085},
  doi          = {10.1109/ASE.2019.00085},
  bibsource    = {dblp computer science bibliography, https://dblp.org}
}

@inproceedings{DBLP:conf/fast/WeiZWLZCSZ21,
  author       = {Junyu Wei and
                  Guangyan Zhang and
                  Yang Wang and
                  Zhiwei Liu and
                  Zhanyang Zhu and
                  Junchao Chen and
                  Tingtao Sun and
                  Qi Zhou},
  editor       = {Marcos K. Aguilera and
                  Gala Yadgar},
  title        = {On the Feasibility of Parser-based Log Compression in Large-Scale
                  Cloud Systems},
  booktitle    = {19th {USENIX} Conference on File and Storage Technologies, {FAST}
                  2021, February 23-25, 2021},
  pages        = {249--262},
  publisher    = {{USENIX} Association},
  year         = {2021},
  url          = {https://www.usenix.org/conference/fast21/presentation/wei},
  bibsource    = {dblp computer science bibliography, https://dblp.org}
}

@inproceedings{DBLP:conf/icse/LiZL024,
  author       = {Xiaoyun Li and
                  Hongyu Zhang and
                  Van{-}Hoang Le and
                  Pengfei Chen},
  title        = {LogShrink: Effective Log Compression by Leveraging Commonality and
                  Variability of Log Data},
  booktitle    = {Proceedings of the 46th {IEEE/ACM} International Conference on Software
                  Engineering, {ICSE} 2024, Lisbon, Portugal, April 14-20, 2024},
  pages        = {23:1--23:12},
  publisher    = {{ACM}},
  year         = {2024},
  url          = {https://doi.org/10.1145/3597503.3608129},
  doi          = {10.1145/3597503.3608129},
  bibsource    = {dblp computer science bibliography, https://dblp.org}
}

@inproceedings{DBLP:conf/kbse/YuW0H24,
  author       = {Siyu Yu and
                  Yifan Wu and
                  Ying Li and
                  Pinjia He},
  editor       = {Vladimir Filkov and
                  Baishakhi Ray and
                  Minghui Zhou},
  title        = {Unlocking the Power of Numbers: Log Compression via Numeric Token
                  Parsing},
  booktitle    = {Proceedings of the 39th {IEEE/ACM} International Conference on Automated
                  Software Engineering, {ASE} 2024, Sacramento, CA, USA, October 27
                  - November 1, 2024},
  pages        = {919--930},
  publisher    = {{ACM}},
  year         = {2024},
  url          = {https://doi.org/10.1145/3691620.3695474},
  doi          = {10.1145/3691620.3695474},
  bibsource    = {dblp computer science bibliography, https://dblp.org}
}

@inproceedings{DBLP:conf/IEEEscc/Chen24,
  author       = {Jiarui Chen},
  title        = {Log Compression via Redundancy Eliminating at Word and Numerical Levels},
  booktitle    = {{IEEE} International Conference on Software Services Engineering,
                  {SSE} 2024, Shenzhen, China, July 7-13, 2024},
  pages        = {115--122},
  publisher    = {{IEEE}},
  year         = {2024},
  url          = {https://doi.org/10.1109/SSE62657.2024.00028},
  doi          = {10.1109/SSE62657.2024.00028},
  bibsource    = {dblp computer science bibliography, https://dblp.org}
}

@inproceedings{DBLP:conf/osdi/RodriguesLY21,
  author       = {Kirk Rodrigues and
                  Yu Luo and
                  Ding Yuan},
  editor       = {Angela Demke Brown and
                  Jay R. Lorch},
  title        = {{CLP:} Efficient and Scalable Search on Compressed Text Logs},
  booktitle    = {15th {USENIX} Symposium on Operating Systems Design and Implementation,
                  {OSDI} 2021, July 14-16, 2021},
  pages        = {183--198},
  publisher    = {{USENIX} Association},
  year         = {2021},
  url          = {https://www.usenix.org/conference/osdi21/presentation/rodrigues},
  bibsource    = {dblp computer science bibliography, https://dblp.org}
}

@inproceedings{DBLP:conf/eurosys/WeiZ00ZSWJ23,
  author       = {Junyu Wei and
                  Guangyan Zhang and
                  Junchao Chen and
                  Yang Wang and
                  Weimin Zheng and
                  Tingtao Sun and
                  Jiesheng Wu and
                  Jiangwei Jiang},
  editor       = {Giuseppe Antonio Di Luna and
                  Leonardo Querzoni and
                  Alexandra Fedorova and
                  Dushyanth Narayanan},
  title        = {LogGrep: Fast and Cheap Cloud Log Storage by Exploiting both Static
                  and Runtime Patterns},
  booktitle    = {Proceedings of the Eighteenth European Conference on Computer Systems,
                  EuroSys 2023, Rome, Italy, May 8-12, 2023},
  pages        = {452--468},
  publisher    = {{ACM}},
  year         = {2023},
  url          = {https://doi.org/10.1145/3552326.3567484},
  doi          = {10.1145/3552326.3567484},
  bibsource    = {dblp computer science bibliography, https://dblp.org}
}

@inproceedings{DBLP:conf/icse/ZhuHLHXZL19,
  author       = {Jieming Zhu and
                  Shilin He and
                  Jinyang Liu and
                  Pinjia He and
                  Qi Xie and
                  Zibin Zheng and
                  Michael R. Lyu},
  editor       = {Helen Sharp and
                  Mike Whalen},
  title        = {Tools and benchmarks for automated log parsing},
  booktitle    = {Proceedings of the 41st International Conference on Software Engineering:
                  Software Engineering in Practice, {ICSE} {(SEIP)} 2019, Montreal,
                  QC, Canada, May 25-31, 2019},
  pages        = {121--130},
  publisher    = {{IEEE} / {ACM}},
  year         = {2019},
  url          = {https://doi.org/10.1109/ICSE-SEIP.2019.00021},
  doi          = {10.1109/ICSE-SEIP.2019.00021},
  bibsource    = {dblp computer science bibliography, https://dblp.org}
}

@article{DBLP:journals/pacmse/ChenPZ25,
  author       = {Zhuangbin Chen and
                  Junsong Pu and
                  Zibin Zheng},
  title        = {Tracezip: Efficient Distributed Tracing via Trace Compression},
  journal      = {Proc. {ACM} Softw. Eng.},
  volume       = {2},
  number       = {{ISSTA}},
  pages        = {411--433},
  year         = {2025},
  url          = {https://doi.org/10.1145/3728888},
  doi          = {10.1145/3728888},
  bibsource    = {dblp computer science bibliography, https://dblp.org}
}

@inproceedings{DBLP:conf/icws/HeZZL17,
  author       = {Pinjia He and
                  Jieming Zhu and
                  Zibin Zheng and
                  Michael R. Lyu},
  editor       = {Ilkay Altintas and
                  Shiping Chen},
  title        = {Drain: An Online Log Parsing Approach with Fixed Depth Tree},
  booktitle    = {2017 {IEEE} International Conference on Web Services, {ICWS} 2017,
                  Honolulu, HI, USA, June 25-30, 2017},
  pages        = {33--40},
  publisher    = {{IEEE}},
  year         = {2017},
  url          = {https://doi.org/10.1109/ICWS.2017.13},
  doi          = {10.1109/ICWS.2017.13},
  bibsource    = {dblp computer science bibliography, https://dblp.org}
}

@inproceedings{DBLP:conf/icse/YuCLWZDZ23,
  author       = {Guangba Yu and
                  Pengfei Chen and
                  Pairui Li and
                  Tianjun Weng and
                  Haibing Zheng and
                  Yuetang Deng and
                  Zibin Zheng},
  title        = {LogReducer: Identify and Reduce Log Hotspots in Kernel on the Fly},
  booktitle    = {45th {IEEE/ACM} International Conference on Software Engineering,
                  {ICSE} 2023, Melbourne, Australia, May 14-20, 2023},
  pages        = {1763--1775},
  publisher    = {{IEEE}},
  year         = {2023},
  url          = {https://doi.org/10.1109/ICSE48619.2023.00151},
  doi          = {10.1109/ICSE48619.2023.00151},
  bibsource    = {dblp computer science bibliography, https://dblp.org}
}

@inproceedings{DBLP:conf/nsdi/ZhangXAVM23,
  author       = {Lei Zhang and
                  Zhiqiang Xie and
                  Vaastav Anand and
                  Ymir Vigfusson and
                  Jonathan Mace},
  editor       = {Mahesh Balakrishnan and
                  Manya Ghobadi},
  title        = {The Benefit of Hindsight: Tracing Edge-Cases in Distributed Systems},
  booktitle    = {20th {USENIX} Symposium on Networked Systems Design and Implementation,
                  {NSDI} 2023, Boston, MA, April 17-19, 2023},
  pages        = {321--339},
  publisher    = {{USENIX} Association},
  year         = {2023},
  url          = {https://www.usenix.org/conference/nsdi23/presentation/zhang-lei},
  bibsource    = {dblp computer science bibliography, https://dblp.org}
}

@inproceedings{DBLP:conf/IEEEcloud/ChenJSLZ24,
  author       = {Zhuangbin Chen and
                  Zhihan Jiang and
                  Yuxin Su and
                  Michael R. Lyu and
                  Zibin Zheng},
  editor       = {Rong N. Chang and
                  Carl K. Chang and
                  Jingwei Yang and
                  Nimanthi L. Atukorala and
                  Zhi Jin and
                  Michael Sheng and
                  Jing Fan and
                  Kenneth Fletcher and
                  Qiang He and
                  Tevfik Kosar and
                  Santonu Sarkar and
                  Sreekrishnan Venkateswaran and
                  Shangguang Wang and
                  Xuanzhe Liu and
                  Seetharami Seelam and
                  Chandra Narayanaswami and
                  Ziliang Zong},
  title        = {Tracemesh: Scalable and Streaming Sampling for Distributed Traces},
  booktitle    = {17th {IEEE} International Conference on Cloud Computing, {CLOUD} 2024,
                  Shenzhen, China, July 7-13, 2024},
  pages        = {54--65},
  publisher    = {{IEEE}},
  year         = {2024},
  url          = {https://doi.org/10.1109/CLOUD62652.2024.00016},
  doi          = {10.1109/CLOUD62652.2024.00016},
  bibsource    = {dblp computer science bibliography, https://dblp.org}
}

@inproceedings{DBLP:conf/qsic/JiangHFH08,
  author       = {Zhen Ming Jiang and
                  Ahmed E. Hassan and
                  Parminder Flora and
                  Gilbert Hamann},
  editor       = {Hong Zhu},
  title        = {Abstracting Execution Logs to Execution Events for Enterprise Applications
                  (Short Paper)},
  booktitle    = {Proceedings of the Eighth International Conference on Quality Software,
                  {QSIC} 2008, 12-13 August 2008, Oxford, {UK}},
  pages        = {181--186},
  publisher    = {{IEEE} Computer Society},
  year         = {2008},
  url          = {https://doi.org/10.1109/QSIC.2008.50},
  doi          = {10.1109/QSIC.2008.50},
  bibsource    = {dblp computer science bibliography, https://dblp.org}
}

@inproceedings{DBLP:conf/kdd/MakanjuZM09,
  author       = {Adetokunbo Makanju and
                  Nur Zincir{-}Heywood and
                  Evangelos E. Milios},
  editor       = {John F. Elder IV and
                  Fran{\c{c}}oise Fogelman{-}Souli{\'{e}} and
                  Peter A. Flach and
                  Mohammed Javeed Zaki},
  title        = {Clustering event logs using iterative partitioning},
  booktitle    = {Proceedings of the 15th {ACM} {SIGKDD} International Conference on
                  Knowledge Discovery and Data Mining, Paris, France, June 28 - July
                  1, 2009},
  pages        = {1255--1264},
  publisher    = {{ACM}},
  year         = {2009},
  url          = {https://doi.org/10.1145/1557019.1557154},
  doi          = {10.1145/1557019.1557154},
  bibsource    = {dblp computer science bibliography, https://dblp.org}
}

@inproceedings{DBLP:conf/msr/NagappanV10,
  author       = {Meiyappan Nagappan and
                  Mladen A. Vouk},
  editor       = {Jim Whitehead and
                  Thomas Zimmermann},
  title        = {Abstracting log lines to log event types for mining software system
                  logs},
  booktitle    = {Proceedings of the 7th International Working Conference on Mining
                  Software Repositories, {MSR} 2010 (Co-located with ICSE), Cape Town,
                  South Africa, May 2-3, 2010, Proceedings},
  pages        = {114--117},
  publisher    = {{IEEE} Computer Society},
  year         = {2010},
  url          = {https://doi.org/10.1109/MSR.2010.5463281},
  doi          = {10.1109/MSR.2010.5463281},
  bibsource    = {dblp computer science bibliography, https://dblp.org}
}

@inproceedings{DBLP:conf/cikm/TangLP11,
  author       = {Liang Tang and
                  Tao Li and
                  Chang{-}Shing Perng},
  editor       = {Craig Macdonald and
                  Iadh Ounis and
                  Ian Ruthven},
  title        = {LogSig: generating system events from raw textual logs},
  booktitle    = {Proceedings of the 20th {ACM} Conference on Information and Knowledge
                  Management, {CIKM} 2011, Glasgow, United Kingdom, October 24-28, 2011},
  pages        = {785--794},
  publisher    = {{ACM}},
  year         = {2011},
  url          = {https://doi.org/10.1145/2063576.2063690},
  doi          = {10.1145/2063576.2063690},
  bibsource    = {dblp computer science bibliography, https://dblp.org}
}

@inproceedings{DBLP:conf/iwpc/MessaoudiPBBS18,
  author       = {Salma Messaoudi and
                  Annibale Panichella and
                  Domenico Bianculli and
                  Lionel C. Briand and
                  Raimondas Sasnauskas},
  editor       = {Foutse Khomh and
                  Chanchal K. Roy and
                  Janet Siegmund},
  title        = {A search-based approach for accurate identification of log message
                  formats},
  booktitle    = {Proceedings of the 26th Conference on Program Comprehension, {ICPC}
                  2018, Gothenburg, Sweden, May 27-28, 2018},
  pages        = {167--177},
  publisher    = {{ACM}},
  year         = {2018},
  url          = {https://doi.org/10.1145/3196321.3196340},
  doi          = {10.1145/3196321.3196340},
  bibsource    = {dblp computer science bibliography, https://dblp.org}
}

@inproceedings{DBLP:conf/IEEEscc/Mizutani13,
  author       = {Masayoshi Mizutani},
  title        = {Incremental Mining of System Log Format},
  booktitle    = {2013 {IEEE} International Conference on Services Computing, Santa
                  Clara, CA, USA, June 28 - July 3, 2013},
  pages        = {595--602},
  publisher    = {{IEEE} Computer Society},
  year         = {2013},
  url          = {https://doi.org/10.1109/SCC.2013.73},
  doi          = {10.1109/SCC.2013.73},
  bibsource    = {dblp computer science bibliography, https://dblp.org}
}

@inproceedings{DBLP:conf/icdm/Du016,
  author       = {Min Du and
                  Feifei Li},
  editor       = {Francesco Bonchi and
                  Josep Domingo{-}Ferrer and
                  Ricardo Baeza{-}Yates and
                  Zhi{-}Hua Zhou and
                  Xindong Wu},
  title        = {Spell: Streaming Parsing of System Event Logs},
  booktitle    = {{IEEE} 16th International Conference on Data Mining, {ICDM} 2016,
                  December 12-15, 2016, Barcelona, Spain},
  pages        = {859--864},
  publisher    = {{IEEE} Computer Society},
  year         = {2016},
  url          = {https://doi.org/10.1109/ICDM.2016.0103},
  doi          = {10.1109/ICDM.2016.0103},
  bibsource    = {dblp computer science bibliography, https://dblp.org}
}

@inproceedings{DBLP:conf/sigmod/ChristensenL13,
  author       = {Robert Christensen and
                  Feifei Li},
  editor       = {Kenneth A. Ross and
                  Divesh Srivastava and
                  Dimitris Papadias},
  title        = {Adaptive log compression for massive log data},
  booktitle    = {Proceedings of the {ACM} {SIGMOD} International Conference on Management
                  of Data, {SIGMOD} 2013, New York, NY, USA, June 22-27, 2013},
  pages        = {1283--1284},
  publisher    = {{ACM}},
  year         = {2013},
  url          = {https://doi.org/10.1145/2463676.2465341},
  doi          = {10.1145/2463676.2465341},
  bibsource    = {dblp computer science bibliography, https://dblp.org}
}

@inproceedings{DBLP:conf/trustcom/FengWL16,
  author       = {Bo Feng and
                  Chentao Wu and
                  Jie Li},
  title        = {{MLC:} An Efficient Multi-level Log Compression Method for Cloud Backup
                  Systems},
  booktitle    = {2016 {IEEE} Trustcom/BigDataSE/ISPA, Tianjin, China, August 23-26,
                  2016},
  pages        = {1358--1365},
  publisher    = {{IEEE}},
  year         = {2016},
  url          = {https://doi.org/10.1109/TrustCom.2016.0215},
  doi          = {10.1109/TRUSTCOM.2016.0215},
  bibsource    = {dblp computer science bibliography, https://dblp.org}
}

@article{DBLP:journals/csur/HeHCYSL21,
  author       = {Shilin He and
                  Pinjia He and
                  Zhuangbin Chen and
                  Tianyi Yang and
                  Yuxin Su and
                  Michael R. Lyu},
  title        = {A Survey on Automated Log Analysis for Reliability Engineering},
  journal      = {{ACM} Comput. Surv.},
  volume       = {54},
  number       = {6},
  pages        = {130:1--130:37},
  year         = {2022},
  url          = {https://doi.org/10.1145/3460345},
  doi          = {10.1145/3460345},
  bibsource    = {dblp computer science bibliography, https://dblp.org}
}

@inproceedings{DBLP:conf/ccs/Du0ZS17,
  author       = {Min Du and
                  Feifei Li and
                  Guineng Zheng and
                  Vivek Srikumar},
  editor       = {Bhavani Thuraisingham and
                  David Evans and
                  Tal Malkin and
                  Dongyan Xu},
  title        = {DeepLog: Anomaly Detection and Diagnosis from System Logs through
                  Deep Learning},
  booktitle    = {Proceedings of the 2017 {ACM} {SIGSAC} Conference on Computer and
                  Communications Security, {CCS} 2017, Dallas, TX, USA, October 30 -
                  November 03, 2017},
  pages        = {1285--1298},
  publisher    = {{ACM}},
  year         = {2017},
  url          = {https://doi.org/10.1145/3133956.3134015},
  doi          = {10.1145/3133956.3134015},
  bibsource    = {dblp computer science bibliography, https://dblp.org}
}

@inproceedings{DBLP:conf/icml/XuHFPJ10,
  author       = {Wei Xu and
                  Ling Huang and
                  Armando Fox and
                  David A. Patterson and
                  Michael I. Jordan},
  editor       = {Johannes F{\"{u}}rnkranz and
                  Thorsten Joachims},
  title        = {Detecting Large-Scale System Problems by Mining Console Logs},
  booktitle    = {Proceedings of the 27th International Conference on Machine Learning
                  (ICML-10), June 21-24, 2010, Haifa, Israel},
  pages        = {37--46},
  publisher    = {Omnipress},
  year         = {2010},
  url          = {https://icml.cc/Conferences/2010/papers/902.pdf},
  bibsource    = {dblp computer science bibliography, https://dblp.org}
}

@inproceedings{DBLP:conf/osdi/ChowMFPW14,
  author       = {Michael Chow and
                  David Meisner and
                  Jason Flinn and
                  Daniel Peek and
                  Thomas F. Wenisch},
  editor       = {Jason Flinn and
                  Hank Levy},
  title        = {The Mystery Machine: End-to-end Performance Analysis of Large-scale
                  Internet Services},
  booktitle    = {11th {USENIX} Symposium on Operating Systems Design and Implementation,
                  {OSDI} '14, Broomfield, CO, USA, October 6-8, 2014},
  pages        = {217--231},
  publisher    = {{USENIX} Association},
  year         = {2014},
  url          = {https://www.usenix.org/conference/osdi14/technical-sessions/presentation/chow},
  bibsource    = {dblp computer science bibliography, https://dblp.org}
}

@inproceedings{DBLP:conf/usenix/YangPO18,
  author       = {Stephen Yang and
                  Seo Jin Park and
                  John K. Ousterhout},
  editor       = {Haryadi S. Gunawi and
                  Benjamin C. Reed},
  title        = {NanoLog: {A} Nanosecond Scale Logging System},
  booktitle    = {Proceedings of the 2018 {USENIX} Annual Technical Conference, {USENIX}
                  {ATC} 2018, Boston, MA, USA, July 11-13, 2018},
  pages        = {335--350},
  publisher    = {{USENIX} Association},
  year         = {2018},
  url          = {https://www.usenix.org/conference/atc18/presentation/yang-stephen},
  bibsource    = {dblp computer science bibliography, https://dblp.org}
}

@article{DBLP:journals/rfc/rfc1951,
  author       = {Peter Deutsch},
  title        = {{DEFLATE} Compressed Data Format Specification version 1.3},
  journal      = {{RFC}},
  volume       = {1951},
  pages        = {1--17},
  year         = {1996},
  url          = {https://doi.org/10.17487/RFC1951},
  doi          = {10.17487/RFC1951},
  bibsource    = {dblp computer science bibliography, https://dblp.org}
}

@article{DBLP:journals/tit/ZivL77,
  author       = {Jacob Ziv and
                  Abraham Lempel},
  title        = {A universal algorithm for sequential data compression},
  journal      = {{IEEE} Trans. Inf. Theory},
  volume       = {23},
  number       = {3},
  pages        = {337--343},
  year         = {1977},
  url          = {https://doi.org/10.1109/TIT.1977.1055714},
  doi          = {10.1109/TIT.1977.1055714},
  bibsource    = {dblp computer science bibliography, https://dblp.org}
}

@article{DBLP:journals/ese/YaoLSH20,
  author       = {Kundi Yao and
                  Heng Li and
                  Weiyi Shang and
                  Ahmed E. Hassan},
  title        = {A study of the performance of general compressors on log files},
  journal      = {Empir. Softw. Eng.},
  volume       = {25},
  number       = {5},
  pages        = {3043--3085},
  year         = {2020},
  url          = {https://doi.org/10.1007/s10664-020-09822-x},
  doi          = {10.1007/S10664-020-09822-X},
  bibsource    = {dblp computer science bibliography, https://dblp.org}
}

@article{DBLP:journals/tit/ZivL78,
  author       = {Jacob Ziv and
                  Abraham Lempel},
  title        = {Compression of individual sequences via variable-rate coding},
  journal      = {{IEEE} Trans. Inf. Theory},
  volume       = {24},
  number       = {5},
  pages        = {530--536},
  year         = {1978},
  url          = {https://doi.org/10.1109/TIT.1978.1055934},
  doi          = {10.1109/TIT.1978.1055934},
  bibsource    = {dblp computer science bibliography, https://dblp.org}
}

@inproceedings{DBLP:conf/icse/Khan0BB22,
  author       = {Zanis Ali Khan and
                  Donghwan Shin and
                  Domenico Bianculli and
                  Lionel C. Briand},
  title        = {Guidelines for Assessing the Accuracy of Log Message Template Identification
                  Techniques},
  booktitle    = {44th {IEEE/ACM} 44th International Conference on Software Engineering,
                  {ICSE} 2022, Pittsburgh, PA, USA, May 25-27, 2022},
  pages        = {1095--1106},
  publisher    = {{ACM}},
  year         = {2022},
  url          = {https://doi.org/10.1145/3510003.3510101},
  doi          = {10.1145/3510003.3510101},
  bibsource    = {dblp computer science bibliography, https://dblp.org}
}

@inproceedings{DBLP:conf/issta/JiangL00HGCZL24,
  author       = {Zhihan Jiang and
                  Jinyang Liu and
                  Junjie Huang and
                  Yichen Li and
                  Yintong Huo and
                  Jiazhen Gu and
                  Zhuangbin Chen and
                  Jieming Zhu and
                  Michael R. Lyu},
  editor       = {Maria Christakis and
                  Michael Pradel},
  title        = {A Large-Scale Evaluation for Log Parsing Techniques: How Far Are We?},
  booktitle    = {Proceedings of the 33rd {ACM} {SIGSOFT} International Symposium on
                  Software Testing and Analysis, {ISSTA} 2024, Vienna, Austria, September
                  16-20, 2024},
  pages        = {223--234},
  publisher    = {{ACM}},
  year         = {2024},
  url          = {https://doi.org/10.1145/3650212.3652123},
  doi          = {10.1145/3650212.3652123},
  bibsource    = {dblp computer science bibliography, https://dblp.org}
}

@inproceedings{DBLP:conf/icse/LiLCSHLZ23,
  author       = {Zhenhao Li and
                  Chuan Luo and
                  Tse{-}Hsun Chen and
                  Weiyi Shang and
                  Shilin He and
                  Qingwei Lin and
                  Dongmei Zhang},
  title        = {Did We Miss Something Important? Studying and Exploring Variable-Aware
                  Log Abstraction},
  booktitle    = {45th {IEEE/ACM} International Conference on Software Engineering,
                  {ICSE} 2023, Melbourne, Australia, May 14-20, 2023},
  pages        = {830--842},
  publisher    = {{IEEE}},
  year         = {2023},
  url          = {https://doi.org/10.1109/ICSE48619.2023.00078},
  doi          = {10.1109/ICSE48619.2023.00078},
  bibsource    = {dblp computer science bibliography, https://dblp.org}
}

@inproceedings{DBLP:conf/sigsoft/FuYXLLZY22,
  author       = {Ying Fu and
                  Meng Yan and
                  Jian Xu and
                  Jianguo Li and
                  Zhongxin Liu and
                  Xiaohong Zhang and
                  Dan Yang},
  editor       = {Abhik Roychoudhury and
                  Cristian Cadar and
                  Miryung Kim},
  title        = {Investigating and improving log parsing in practice},
  booktitle    = {Proceedings of the 30th {ACM} Joint European Software Engineering
                  Conference and Symposium on the Foundations of Software Engineering,
                  {ESEC/FSE} 2022, Singapore, Singapore, November 14-18, 2022},
  pages        = {1566--1577},
  publisher    = {{ACM}},
  year         = {2022},
  url          = {https://doi.org/10.1145/3540250.3558947},
  doi          = {10.1145/3540250.3558947},
  bibsource    = {dblp computer science bibliography, https://dblp.org}
}

@article{DBLP:journals/corr/abs-2107-05908,
  author       = {Zhuangbin Chen and
                  Jinyang Liu and
                  Wenwei Gu and
                  Yuxin Su and
                  Michael R. Lyu},
  title        = {Experience Report: Deep Learning-based System Log Analysis for Anomaly
                  Detection},
  journal      = {CoRR},
  volume       = {abs/2107.05908},
  year         = {2021},
  url          = {https://arxiv.org/abs/2107.05908},
  eprinttype    = {arXiv},
  eprint       = {2107.05908},
  bibsource    = {dblp computer science bibliography, https://dblp.org}
}

@inproceedings{DBLP:conf/sigsoft/ChenKLZZXZYSXDG20,
  author       = {Zhuangbin Chen and
                  Yu Kang and
                  Liqun Li and
                  Xu Zhang and
                  Hongyu Zhang and
                  Hui Xu and
                  Yangfan Zhou and
                  Li Yang and
                  Jeffrey Sun and
                  Zhangwei Xu and
                  Yingnong Dang and
                  Feng Gao and
                  Pu Zhao and
                  Bo Qiao and
                  Qingwei Lin and
                  Dongmei Zhang and
                  Michael R. Lyu},
  editor       = {Prem Devanbu and
                  Myra B. Cohen and
                  Thomas Zimmermann},
  title        = {Towards intelligent incident management: why we need it and how we
                  make it},
  booktitle    = {{ESEC/FSE} '20: 28th {ACM} Joint European Software Engineering Conference
                  and Symposium on the Foundations of Software Engineering, Virtual
                  Event, USA, November 8-13, 2020},
  pages        = {1487--1497},
  publisher    = {{ACM}},
  year         = {2020},
  url          = {https://doi.org/10.1145/3368089.3417055},
  doi          = {10.1145/3368089.3417055},
  bibsource    = {dblp computer science bibliography, https://dblp.org}
}

@inproceedings{DBLP:conf/osdi/ZhaoRLYS16,
  author       = {Xu Zhao and
                  Kirk Rodrigues and
                  Yu Luo and
                  Ding Yuan and
                  Michael Stumm},
  editor       = {Kimberly Keeton and
                  Timothy Roscoe},
  title        = {Non-Intrusive Performance Profiling for Entire Software Stacks Based
                  on the Flow Reconstruction Principle},
  booktitle    = {12th {USENIX} Symposium on Operating Systems Design and Implementation,
                  {OSDI} 2016, Savannah, GA, USA, November 2-4, 2016},
  pages        = {603--618},
  publisher    = {{USENIX} Association},
  year         = {2016},
  url          = {https://www.usenix.org/conference/osdi16/technical-sessions/presentation/zhao},
  bibsource    = {dblp computer science bibliography, https://dblp.org}
}

@inproceedings{DBLP:conf/asplos/YuanMXTZP10,
  author       = {Ding Yuan and
                  Haohui Mai and
                  Weiwei Xiong and
                  Lin Tan and
                  Yuanyuan Zhou and
                  Shankar Pasupathy},
  editor       = {James C. Hoe and
                  Vikram S. Adve},
  title        = {SherLog: error diagnosis by connecting clues from run-time logs},
  booktitle    = {Proceedings of the 15th International Conference on Architectural
                  Support for Programming Languages and Operating Systems, {ASPLOS}
                  2010, Pittsburgh, Pennsylvania, USA, March 13-17, 2010},
  pages        = {143--154},
  publisher    = {{ACM}},
  year         = {2010},
  url          = {https://doi.org/10.1145/1736020.1736038},
  doi          = {10.1145/1736020.1736038},
  bibsource    = {dblp computer science bibliography, https://dblp.org}
}

@inproceedings{DBLP:conf/usenix/DingZLZLFZX15,
  author       = {Rui Ding and
                  Hucheng Zhou and
                  Jian{-}Guang Lou and
                  Hongyu Zhang and
                  Qingwei Lin and
                  Qiang Fu and
                  Dongmei Zhang and
                  Tao Xie},
  editor       = {Shan Lu and
                  Erik Riedel},
  title        = {Log2: {A} Cost-Aware Logging Mechanism for Performance Diagnosis},
  booktitle    = {Proceedings of the 2015 {USENIX} Annual Technical Conference, {USENIX}
                  {ATC} 2015, July 8-10, Santa Clara, CA, {USA}},
  pages        = {139--150},
  publisher    = {{USENIX} Association},
  year         = {2015},
  url          = {https://www.usenix.org/conference/atc15/technical-session/presentation/ding},
  bibsource    = {dblp computer science bibliography, https://dblp.org}
}

@article{DBLP:journals/ese/KhanSBB24,
  author       = {Zanis Ali Khan and
                  Donghwan Shin and
                  Domenico Bianculli and
                  Lionel C. Briand},
  title        = {Impact of log parsing on deep learning-based anomaly detection},
  journal      = {Empir. Softw. Eng.},
  volume       = {29},
  number       = {6},
  pages        = {139},
  year         = {2024},
  url          = {https://doi.org/10.1007/s10664-024-10533-w},
  doi          = {10.1007/S10664-024-10533-W},
  bibsource    = {dblp computer science bibliography, https://dblp.org}
}

@article{DBLP:journals/infsof/YuanLSL20,
  author       = {Wei Yuan and
                  Shan Lu and
                  Hailong Sun and
                  Xudong Liu},
  title        = {How are distributed bugs diagnosed and fixed through system logs?},
  journal      = {Inf. Softw. Technol.},
  volume       = {119},
  year         = {2020},
  url          = {https://doi.org/10.1016/j.infsof.2019.106234},
  doi          = {10.1016/J.INFSOF.2019.106234},
  bibsource    = {dblp computer science bibliography, https://dblp.org}
}

@inproceedings{DBLP:conf/issre/HeZHL16,
  author       = {Shilin He and
                  Jieming Zhu and
                  Pinjia He and
                  Michael R. Lyu},
  title        = {Experience Report: System Log Analysis for Anomaly Detection},
  booktitle    = {27th {IEEE} International Symposium on Software Reliability Engineering,
                  {ISSRE} 2016, Ottawa, ON, Canada, October 23-27, 2016},
  pages        = {207--218},
  publisher    = {{IEEE} Computer Society},
  year         = {2016},
  url          = {https://doi.org/10.1109/ISSRE.2016.21},
  doi          = {10.1109/ISSRE.2016.21},
  bibsource    = {dblp computer science bibliography, https://dblp.org}
}

@inproceedings{DBLP:conf/osdi/WangGR0ZWFC024,
  author       = {Rui Wang and
                  Devin Gibson and
                  Kirk Rodrigues and
                  Yu Luo and
                  Yun Zhang and
                  Kaibo Wang and
                  Yupeng Fu and
                  Ting Chen and
                  Ding Yuan},
  editor       = {Ada Gavrilovska and
                  Douglas B. Terry},
  title        = {{\(\mu\)}Slope: High Compression and Fast Search on Semi-Structured
                  Logs},
  booktitle    = {18th {USENIX} Symposium on Operating Systems Design and Implementation,
                  {OSDI} 2024, Santa Clara, CA, USA, July 10-12, 2024},
  pages        = {529--544},
  publisher    = {{USENIX} Association},
  year         = {2024},
  url          = {https://www.usenix.org/conference/osdi24/presentation/wang-rui},
  bibsource    = {dblp computer science bibliography, https://dblp.org}
}

@article{DBLP:journals/corr/abs-2509-26463,
  author       = {Junsong Pu and
                  Yichen Li and
                  Zhuangbin Chen and
                  Jinyang Liu and
                  Zhihan Jiang and
                  Jianjun Chen and
                  Rui Shi and
                  Zibin Zheng and
                  Tieying Zhang},
  title        = {ErrorPrism: Reconstructing Error Propagation Paths in Cloud Service
                  Systems},
  journal      = {CoRR},
  volume       = {abs/2509.26463},
  year         = {2025},
  url          = {https://doi.org/10.48550/arXiv.2509.26463},
  doi          = {10.48550/ARXIV.2509.26463},
  eprinttype    = {arXiv},
  eprint       = {2509.26463},
  bibsource    = {dblp computer science bibliography, https://dblp.org}
}

@inproceedings{DBLP:conf/sigsoft/Zhou0X0JLXH19,
  author       = {Xiang Zhou and
                  Xin Peng and
                  Tao Xie and
                  Jun Sun and
                  Chao Ji and
                  Dewei Liu and
                  Qilin Xiang and
                  Chuan He},
  editor       = {Marlon Dumas and
                  Dietmar Pfahl and
                  Sven Apel and
                  Alessandra Russo},
  title        = {Latent error prediction and fault localization for microservice applications
                  by learning from system trace logs},
  booktitle    = {Proceedings of the {ACM} Joint Meeting on European Software Engineering
                  Conference and Symposium on the Foundations of Software Engineering,
                  {ESEC/SIGSOFT} {FSE} 2019, Tallinn, Estonia, August 26-30, 2019},
  pages        = {683--694},
  publisher    = {{ACM}},
  year         = {2019},
  url          = {https://doi.org/10.1145/3338906.3338961},
  doi          = {10.1145/3338906.3338961},
  bibsource    = {dblp computer science bibliography, https://dblp.org}
}


\end{document}